\documentstyle[graphicx]{mn2e}
\footnotesize
\newdimen\minuswidth    
\setbox0=\hbox{$-$}
\minuswidth=\wd0
\catcode`@=\active
\def@{\kern\minuswidth}
\newdimen\digitwidth    
\setbox0=\hbox{\rm0}
\digitwidth=\wd0
\catcode`!=\active
\def!{\kern\digitwidth}
\normalsize
\title[A statistical study of 233 pulsar proper motions]
{A statistical study of 233 pulsar proper motions}
\author[G. Hobbs et al.]
{G. Hobbs,$^{1,2}$
D.R. Lorimer,$^2$
A.G. Lyne,$^2$
M. Kramer$^2$
\\
$^1$ Australia Telescope National Facility, CSIRO, PO~Box~76, Epping
NSW~1710, Australia \\
$^2$ University of Manchester, Jodrell Bank Observatory, Macclesfield,
Cheshire SK11~9DL \\}

\let\leqslant=\leq 
\date{}
\begin{document}
\maketitle
\newcommand{\setthebls}{
}
\setthebls
\begin{abstract}
We present and analyse a catalogue of 233 pulsars with proper motion
measurements. The sample contains a wide variety of pulsars including
recycled objects and those associated with globular clusters or
supernova remnants. After taking the most precise proper motions for
those pulsars for which multiple measurements are available, the
majority of the proper motions (58\%) are derived from pulsar timing
methods, 41\% using interferometers and the remaining 1\% using
optical telescopes.  Many of the 1-D and 2-D speeds (referring to
speeds measured in one coordinate only and the magnitudes of the
transverse velocities respectively) derived from these measurements
are somewhat lower than earlier estimates because of the use of the
most recent electron density model in determining pulsar distances.
The mean 1-D speeds for the normal and recycled pulsars are 152(10)
and 54(6)\,km\,s$^{-1}$ respectively.  The corresponding mean 2-D
speeds are 246(22) and 87(13)\,km\,s$^{-1}$.  PSRs~B2011+38 and
B2224+64 have the highest inferred 2-D speeds of $\sim
1600$\,km\,s$^{-1}$.  We study the mean speeds for different
subsamples and find that, in general, they agree with previous
results.  Applying a novel deconvolution technique to the sample of 73
pulsars with characteristic ages less than 3\,Myr, we find the mean
3-D pulsar birth velocity to be 400(40) km\,s$^{-1}$. The distribution
of velocities is well described by a Maxwellian distribution with 1-D
rms $\sigma=265$~km~s$^{-1}$.  There is no evidence for a bimodal
velocity distribution.  The proper motions for PSRs~B1830$-$08 and
B2334$+$61 are consistent with their proposed associations with the
supernova remnants W41 and G114.3+0.3 respectively.
\end{abstract}

\section{Introduction}

Neutron stars are high-velocity objects. From observations of radio
pulsars, it has long been known (Gunn \& Ostriker \nocite{go70}1970)
that as a population they are generally moving much faster than their
presumed progenitor population, the massive OB stars.  While the
astrophysical applications of a large sample of pulsar velocities are
manyfold, most importantly the physical mechanism for the high
velocities is not well understood. Lai, Chernoff \& Cordes
(2001)\nocite{lcc01} have summarised the various discussed causes for
high pulsar velocities.  These mechanisms, in general, only predict a
limited range of velocities, so that a large sample of pulsar
velocities can in principle be used to test the mechanisms.

 \begin{figure*}
  \includegraphics[width=10cm,angle=-90]{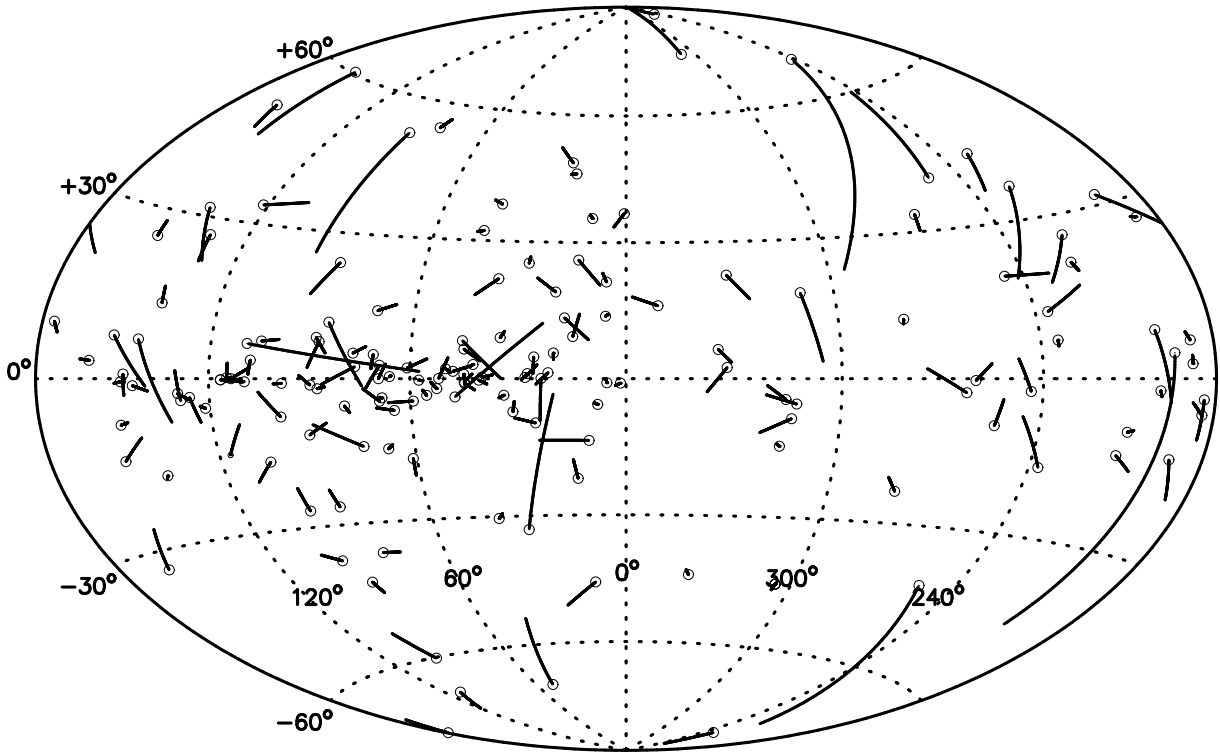}
  \caption{The Galactic motions of the pulsars in our sample.  A
  pulsar is currently at the position indicated by a circle and the
  track indicates its motion for the last 1\,Myr assuming no radial
  velocity.}\label{fg:aitoff}
 \end{figure*}

Pulsar velocities are determined from measurements of their proper
motion and distance.  Prior to the recent publication by Hobbs et
al.~(2004; hereafter Paper~I)\nocite{hlk+04} approximately 138 pulsar
proper motions had been measured, with 68 having values greater than
2$\sigma$.  Most millisecond pulsar proper motions had been obtained
using timing methods (e.g. Nice \& Taylor 1995\nocite{nt95}; Toscano
et al.~1999b\nocite{tsb+99}; Wolszczan et al.~2000\nocite{wdk+00} and
Freire et al.~2003\nocite{fck+03}).  The poor rotational stability
observed for young pulsars\footnote{We use the words `young' and `old'
throughout this paper to refer to pulsars with small and large
characteristic ages, $\tau_c = P/2\dot{P}$.  The characteristic age
assumes that the current rotational period of the pulsar is much
greater than the initial period and that pulsar slow-down is due to a
constant magnetic dipole field.  The characteristic age may not be a
good estimation of the true age if dipole braking is not strictly
followed.}  (referred to as timing noise) affected early measurements
of proper motions using timing methods; for these, interferometers had
been used to obtain the proper motions (for example, Lyne, Anderson \&
Salter 1982; Bailes et al.~1990\nocite{bmk+90b}; Harrison, Lyne \&
Anderson 1993; Fomalont et al.~1997 and Brisken et al.~2002).
\nocite{hla93} \nocite{bbgt02} \nocite{fgml97} \nocite{las82} In
Paper~I we described a new method, based on the modelling and removal
of timing noise from the timing residuals of young pulsars using
harmonically related sinusoids, that allowed accurate proper motions
to be determined using standard timing methods.  These proper motions
were shown to be consistent with proper motion measurements obtained
using interferometers. We obtained proper motions for 301 pulsars with
164 having values greater than 2$\sigma$ or with an uncertainty less
than 15\,mas\,yr$^{-1}$ in at least one coordinate.  For 87 of these
pulsars, these provide more precise proper motions than the earlier
measurements.

Direct distance measurements have only been obtained for a small
number of pulsars. The majority of distances are estimated from the
dispersion measure and a Galactic electron density model.  Using the
Taylor \& Cordes (1993; hereafter TC93)\nocite{tc93} model, Lyne \&
Lorimer (1994)\nocite{ll94} found the mean pulsar birth
velocity\footnote{Throughout this paper, we will present the
uncertainties on parameters as a value in parenthesis after each
quantity.  The value represents the error (at the 68\% confidence
level) in the least significant digit.} to be 450(90)\,km\,s$^{-1}$.
Recently, Cordes \& Lazio (2002; hereafter CL02)\nocite{cl02}
provided an updated model which, on average, predicts somewhat smaller
distances than TC93 which will clearly have an impact on the
calculated velocities.  Hereafter, we designate the velocities derived
from the two models as $V^{TC}$ and $V^{CL}$.

Clearly, with these new proper motion determinations and a new
electron density model it is productive to revisit the statistics of
pulsar velocities.  In Section 2, we describe the sample of proper
motions used in this work which combines new results published in
Paper~I with other proper motion values in the literature.  In Section
3, we highlight the effect of using the CL02 electron density model to
obtain pulsar distances and velocities. In Section 4 we examine the
statistical properties of various sub-samples of the observed sample.
Following a brief discussion of the motion of pulsars in the
Galactic plane in Section 5, we estimate the 3-D birth speed
distribution of non-millisecond pulsars in Section 6. Finally, in
Section 7, we summarise our main results and conclusions.

\section{The proper motion sample}

 Proper motions were selected from the literature with more precise
 measurements taking precedence if multiple measurements exist for a
 specific pulsar. In practice, these proper motions were obtained from
 the ATNF pulsar
 catalogue\footnote{http://www.atnf.csiro.au/research/pulsar/psrcat}
 (Manchester et al.~2005)\nocite{mhth05} or from Paper~I.  These
 proper motion measurements all have values greater than 2$\sigma$ or
 have an uncertainty less than 15\,mas\,yr$^{-1}$. As discussed in
 Paper~I, for a pulsar lying in the ecliptic plane it is not possible,
 using timing methods, to obtain a precise estimate of its proper
 motion in ecliptic latitude. Therefore, only the longitudinal
 component of the proper motion has been measured for many of the
 pulsars in the sample. The timing solutions for the 87 pulsars
 obtained from Paper~I have been updated using the most recent data
 available from the Jodrell Bank Observatory data archive.  The timing
 solutions were obtained in an identical manner to that described in
 Paper~I.  The most recent observations used in these timing solutions
 are from February 2004.

 \begin{table*}\caption{Pulsar proper motions.  For each pulsar, the
two components of its proper motion are provided followed by a flag
stating whether these are given in celestial (C) or ecliptic (E)
coordinates. The remaining columns provide the total proper motion, a
reference for the measurement (see the end of the table for the full
bibliographic reference), whether the proper motion was determined
using interferometric (I), timing (T) or optical (O) means, the
distance according to the CL02 distance model or from independent
distance estimates if available, the magnitude of the transverse
velocity assuming the CL02 or independent distance estimates (ind.) and the
difference $\Delta V_T=V_T^{TC}-V_T^{CL}$ between the velocities as measured assuming the CL02 and
TC93 distances.}\label{tb:pmsample}\begin{tabular}{llllllllrr}\hline
PSR & $\mu_{\rm long}$ & $\mu_{\rm lat}$ & Coord. & $\mu_{\rm tot}$ & Ref. & Obs. & Dist. & $V_T^{CL}$ & $\Delta V_T$ \\
 & (mas yr$^{-1}$) & (mas yr$^{-1}$) & & (mas yr$^{-1}$) & & Type &
 (kpc) & (km\,s$^{-1}$) & (km\,s$^{-1}$) \\  
\hline
B0011+47     & 19.3(18)     & $-$19.7(15)   & C  & 27.6(17)    & 1   & I  & 1.56      & 204   & 34    \\
B0021$-$72C    & 5.2(5)       & $-$3.2(4)     & C  & 6.1(5)      & 2   & T  & 5.00      & 145   & ind.     \\
B0021$-$72D    & 3.8(4)       & $-$2.4(4)     & C  & 4.5(4)      & 2   & T  & 5.00      & 107   & ind.     \\
B0021$-$72E    & 6.2(5)       & $-$2.7(5)     & C  & 6.8(5)      & 2   & T  & 5.00      & 161   & ind.     \\
B0021$-$72F    & 4.6(5)       & $-$2.9(6)     & C  & 5.4(6)      & 2   & T  & 5.00      & 128   & ind.     \\ \\

B0021$-$72G    & 4.2(14)      & $-$3.7(12)    & C  & 5.6(14)     & 2   & T  & 5.00      & 133   & ind.     \\
B0021$-$72H    & 5.1(12)      & $-$3.6(12)    & C  & 6.2(12)     & 2   & T  & 5.00      & 147   & ind.     \\
B0021$-$72I    & 4.9(16)      & $-$3.9(19)    & C  & 6.3(18)     & 2   & T  & 5.00      & 149   & ind.     \\
B0021$-$72J    & 5.35(13)     & $-$3.42(15)   & C  & 6.35(14)    & 2   & T  & 5.00      & 150   & ind.     \\
B0021$-$72N    & 6.8(13)      & $-$1.1(14)    & C  & 6.9(14)     & 2   & T  & 5.00      & 164   & ind.     \\ \\

J0024$-$7204O  & 4.6(11)      & $-$2.9(9)     & C  & 5.4(11)     & 2   & T  & 5.00      & 128   & ind.     \\
J0024$-$7204U  & 5.1(7)       & $-$4.5(7)     & C  & 6.8(7)      & 2   & T  & 5.00      & 161   & ind.     \\
J0034$-$0534   & 5.5(15)      & $-$31(9)      & E  & 31(9)       & 3   & T  & 0.54      & 79    & 65    \\
B0031$-$07     & $-$8(12)       & ---         & E  & ---         & 3   & T  & 0.41      & ---      & ---      \\
B0037+56     & 3(10)        & $-$7(13)      & E  & 8(13)       & 3   & T  & 3.05      & 116   & 54    \\ \\

B0114+58     & 7(5)         & 17(6)       & E  & 18(6)       & 3   & T  & 2.23      & 190   & $-$8    \\
J0134$-$2937   & 8(3)         & $-$21(5)      & E  & 22(5)       & 3   & T  & 0.56      & 58    & 127   \\
B0136+57     & $-$11(5)       & $-$19(5)      & C  & 22(5)       & 4    & I  & 2.88      & 300   & 1     \\
B0138+59     & 7(12)        & ---         & E  & ---         & 3   & T  & 2.19      & ---      & ---      \\
B0144+59     & $-$3.8(19)     & 10(3)       & E  & 11(3)       & 3   & T  & 2.22      & 116   & $-$16   \\ \\

B0149$-$16     & 3.1(12)      & $-$27(2)      & C  & 27(2)    & 1   & I  & 0.51      & 66    & 36    \\
B0203$-$40     & ---          & 75(35)      & C  & ---         & 5    & T  & 0.58      & ---      & ---      \\
J0218+4232   & 3(3)         & $-$4(7)       & E  & 5(6)        & 3   & T  & 2.67      & 63    & 75    \\
B0254$-$53     & ---          & 70(15)      & C  & ---         & 5    & T  & 0.73      & ---      & ---      \\
B0301+19     & 6(7)         & $-$37(4)      & C  & 37(5)       & 6    & I  & 0.62      & 109   & 58    \\ \\

B0320+39     & 16(6)        & $-$30(5)      & C  & 34(6)       & 4    & I  & 1.01      & 163   & 77    \\
B0329+54     & 17.0(3)      & $-$9.5(4)     & C  & 19.5(4)     & 7   & I  & 1.06      & 98    & ind.     \\
B0331+45     & $-$4(3)        & 5(8)        & E  & 6(7)        & 3   & T  & 1.64      & 47    & 13    \\
B0353+52     & 6(3)         & $-$9(6)       & E  & 11(6)       & 3   & T  & 2.78      & 145   & 102   \\
B0355+54     & 9.20(18)     & 8.2(4)      & C  & 12.3(3)     & 8   & I  & 1.10      & 64    & 57    \\ \\

B0402+61     & 23(7)        & 61(12)      & E  & 65(12)      & 3   & T  & 2.12      & 653   & 287   \\
B0410+69     & $-$4(5)        & $-$12(7)      & E  & 13(7)       & 3   & T  & 1.18      & 73    & 24    \\
J0437$-$4715   & 121.457(9)   & $-$71.449(9)  & C  & 140.914(9)  & 9   & T  & 0.14      & 94    & ind.     \\
B0447$-$12     & 10(11)       & ---         & E  & ---         & 3   & T  & 1.89      & ---      & ---      \\
B0450$-$18     & 12(8)        & ---         & C  & ---         & 10   & I  & 2.36      & ---      & ---      \\ \\

B0450+55     & 52(6)        & $-$17(2)      & C  & 55(6)       & 4    & I  & 0.67      & 175   & 31    \\
B0458+46     & $-$8(3)        & 8(5)        & C  & 11(5)       & 4    & I  & 1.39      & 72    & 20    \\
B0523+11     & 30(7)        & $-$4(5)       & C  & 30(7)       & 4    & I  & 3.11      & 442   & 650   \\
B0525+21     & ---          & 7(9)        & C  & ---         & 4    & I  & 1.61      & ---      & ---      \\
B0531+21     & $-$13(2)       & 7(3)        & C  & 15(3)       & 11     & O  & 2.00      & 142   & ind.     \\ \\

J0538+2817   & $-$30(6)       & ---         & E  & ---         & 3   & T  & 1.23      & ---      & ---      \\
B0540+23     & 19(7)        & 12(8)       & C  & 22(8)       & 4    & I  & 2.06      & 215   & 154   \\
B0559$-$05     & 18(8)        & $-$16(7)      & C  & 24(8)       & 4    & I  & 3.93      & 447   & 411   \\
B0609+37     & 5(3)         & 5(12)       & E  & 7(9)        & 3   & T  & 0.87      & 29    & 21    \\
J0613$-$0200   & 2.0(4)       & $-$7.0(10)    & C  & 7.3(10)     & 12   & T  & 1.71      & 59    & 17    \\ \\

B0611+22     & $-$4(5)        & $-$3(7)       & C  & 5(6)        & 4    & I  & 2.08      & 49    & 63    \\
J0621+1002   & 3.5(3)       & $-$0.3(9)     & C  & 3.5(4)      & 13   & T  & 1.36      & 23    & 9     \\
B0621$-$04     & 0(10)      & ---         & E  & ---         & 3   & T  & 2.78      & ---      & ---      \\
B0626+24     & $-$7(12)       & 2(12)       & C  & 7(12)       & 4    & I  & 2.24      & 74    & 81    \\
B0628$-$28     & $-$44.6(9)     & 20(3)       & C  & 48.7(13)    & 1  & I  & 1.45      & 335   & 162  \\

\hline \end{tabular}\end{table*}\begin{table*}
\begin{tabular}{llllllllrr}\hline
PSR & $\mu_{\rm long}$ & $\mu_{\rm lat}$ & Coord. & $\mu_{\rm tot}$ & Ref. & Obs. & Dist. & $V_T^{CL}$ & $\Delta V_T$ \\
 & (mas yr$^{-1}$) & (mas yr$^{-1}$) & & (mas yr$^{-1}$) & & Type & (kpc) & (km\,s$^{-1}$) & (km\,s$^{-1}$)\\  
\hline
J0633+1746   & 138(4)       & 97(4)       & C  & 169(4)      & 14   & O  & 0.16      & 128   & ind.     \\
B0643+80     & 19(3)        & $-$1(3)    & C  & 19(3)       & 4    & I  & 1.55      & 140   & 164   \\
B0656+14     & 44.1(6)      & $-$2.4(3)       & C  & 44.2(6)  & 15    & I  & 0.29      & 59    & ind.     \\
B0655+64     & $-$6(5)        & $-$8(8)       & E  & 10(8)       & 3   & T  & 0.49      & 23    & 0    \\
J0711$-$6830   & $-$15.7(5)     & 15.3(6)     & C  & 21.9(6)     & 12   & T  & 0.86      & 90    & 19    \\ \\

B0736$-$40     & $-$14.0(12)    & $-$13(3)      & C  & 19.0(17)    & 1  & I  & 2.64      & 238  & 756   \\
B0740$-$28     & $-$29(2)       & 4(2)        & C  & 29.3(2)     & 10   & I  & 2.07      & 287  & $-$25   \\
J0751+1807   & $-$0.2(10)     & ---         & E  & ---         & 3   & T  & 1.15      & ---      & ---      \\
B0751+32     & $-$4(5)        & 7(3)        & C  & 8(4)        & 4    & I  & 1.52      & 58    & 91    \\
B0756$-$15     & 1(4)      & 4(6)        & C  & 4(6)        & 1  & I  & 2.95      & 56    & 15    \\ \\

B0809+74     & 24.02(9)     & $-$44.0(4)    & C  & 50.1(4)     & 7   & I  & 0.43      & 102   & ind.     \\
B0818$-$13     & 9(9)         & $-$47(6)      & C  & 48(7)       & 10   & I  & 1.99      & 453   & 105   \\
B0820+02     & 5(11)        & $-$1(8)    & C  & 5(11)       & 4    & I  & 1.01      & 24    & 10    \\
B0823+26     & 61(3)        & $-$90(2)      & C  & 109(3)      & 6    & I  & 0.36      & 186   & ind.     \\
B0833$-$45     & $-$49.60(6)    & 29.80(10)   & C  & 57.86(8)    & 16   & I  & 0.29      & 80    & ind.     \\ \\

B0834+06     & 2(5)         & 51(3)       & C  & 51(4)       & 6    & I  & 0.65      & 157   & 17    \\
B0835$-$41     & $-$2.3(18)     & $-$18(4)      & C  & 18(4)       & 1  & I  & 1.60      & 137   & 225   \\
B0906$-$17     & 27(11)       & $-$40(11)     & C  & 48(11)      & 4    & I  & 0.91      & 207   & $-$64  \\
B0919+06     & 18.35(6)     & 86.56(12)   & C  & 88.48(12)   & 17   & I  & 1.20      & 503   & ind.     \\
B0940+16     & ---          & 9(11)       & C  & ---         & 4    & I  & 0.81      & ---      & ---      \\ \\

B0942$-$13     & ---          & $-$22(14)     & C  & ---         & 4    & I  & 0.67      & ---      & ---      \\
B0943+10     & ---          & $-$21(12)     & C  & ---         & 6    & I  & 0.64      & ---      & ---      \\
B0950+08     & $-$2.09(8)     & 29.46(7)    & C  & 29.53(8)    & 7   & I  & 0.26      & 36    & ind.     \\
J1012+5307   & 2.4(2)       & $-$25.2(2)    & C  & 25.3(2)     & 18   & T  & 0.41      & 49    & 13    \\
J1022+1001   & $-$16.3(6)     & ---         & E  & ---         & 3   & T  & 0.45      & ---      & ---      \\ \\

J1024$-$0719   & $-$41(2)       & $-$70(3)      & C  & 81(3)       & 12   & T  & 0.39      & 150   & $-$15   \\
B1039$-$19     & $-$1(3)     & 14(5)       & C  & 14(5)       & 1  & I  & 1.48      & 98    & 113   \\
J1045$-$4509   & $-$5(2)        & 6.0(10)     & C  & 7.8(15)     & 12   & T  & 1.96      & 72    & 47    \\
B1112+50     & 22(3)        & $-$51(3)      & C  & 56(3)       & 4    & I  & 0.45      & 119   & 24    \\
B1114$-$41     & $-$1(5)     & ---         & C  & ---         & 1  & I  & 1.47      & ---      & ---      \\ \\

B1133+16     & $-$74.0(4)     & 368.1(3)    & C  & 375.5(4)    & 7   & I  & 0.36      & 641   & ind.     \\
B1237+25     & $-$106.82(17)  & 49.92(18)   & C  & 117.91(18)  & 7   & I  & 0.86      & 481   & ind.     \\
B1254$-$10     & $-$6(8)        & ---         & E  & ---         & 3   & T  & 1.55      & ---      & ---      \\
B1257+12     & 46.40(10)    & $-$82.2(2)    & C  & 94.39(19)   & 19   & T  & 0.77      & 345   & ind.     \\
B1309$-$12     & $-$1(6)     & ---         & E  & ---         & 3   & T  & 3.14      & ---      & ---      \\ \\

B1322+83     & $-$53(20)      & 13(7)       & C  & 55(20)      & 4    & I  & 0.76      & 198   & 3     \\
B1325$-$43     & 3(7)         & 54(23)      & C  & 54(23)      & 1  & I  & 1.34      & 343   & 243   \\
B1426$-$66     & $-$31(5)       & $-$21(3)      & C  & 37(5)       & 20  & I  & 1.00      & 175   & 140   \\
B1449$-$64     & $-$16.0(10)    & $-$21.3(8)    & C  & 26.6(9)     & 20  & I  & 2.08      & 262   & $-$30   \\
J1455$-$3330   & 5(6)         & 24(12)      & C  & 25(12)      & 12   & T  & 0.53      & 63    & 25    \\ \\

B1451$-$68     & $-$39.5(4)     & $-$12.3(3)    & C  & 41.4(4)     & 20  & I  & 0.45      & 88    & ind.     \\
B1508+55     & $-$70.6(16)    & $-$68.8(12) & C  & 98.6(15)    & 1  & I  & 0.99      & 463  & 444   \\
J1518+4904   & 4.6(6)       & $-$8.1(6)     & E  & 9.3(6)      & 3   & T  & 0.63      & 28    & 3    \\
B1534+12     & 1.340(10)    & $-$25.05(2)   & C  & 25.09(2)  & 21    & T  & 0.95      & 113   & $-$6    \\
B1541+09     & $-$7.3(10)     & $-$4.0(10)    & C  & 8.3(10)     & 1  & I  & 3.49      & 137   & $-$41   \\ \\

B1540$-$06     & $-$17(3)       & $-$4(3)       & C  & 18(3)       & 1  & I  & 0.72      & 61    & 37    \\
B1552$-$31     & $-$9(8)        & ---         & E  & ---         & 3   & T  & 2.33      & ---      & ---      \\
B1556$-$44     & 1(6)      & 14(11)      & C  & 14(11)      & 10   & I  & 2.29      & 152   & $-$47   \\
J1603$-$2531   & $-$10(4)       & ---         & E  & ---         & 3   & T  & 1.87      & ---      & ---      \\
B1600$-$27     & $-$21(11)      & ---         & E  & ---         & 3   & T  & 1.56      & ---      & ---      \\

\hline \end{tabular}\end{table*}\begin{table*}
\begin{tabular}{llllllllrr}\hline
PSR & $\mu_{\rm long}$ & $\mu_{\rm lat}$ & Coord. & $\mu_{\rm tot}$ & Ref. & Obs. & Dist. & $V_T^{CL}$ & $\Delta V_T$ \\
 & (mas yr$^{-1}$) & (mas yr$^{-1}$) & & (mas yr$^{-1}$) & & Type & (kpc) & (km\,s$^{-1}$) & (km\,s$^{-1}$)\\  
\hline
J1603$-$7202   & $-$3.5(3)      & $-$7.8(5)     & C  & 8.5(5)      & 12   & T  & 1.17      & 47    & 19    \\
B1600$-$49     & $-$30(7)       & $-$1(3)    & C  & 30(7)       & 20  & I  & 5.10      & 725   & $-$215  \\
B1604$-$00     & $-$1(14)    & $-$7(9)       & C  & 7(10)       & 6    & I  & 0.67      & 22    & $-$3   \\
B1620$-$09     & $-$17(10)      & ---         & E  & ---         & 3   & T  & $>$50.00  & ---      & ---      \\
B1620$-$26     & $-$13.4(10)    & $-$25(5)      & C  & 28(5)       & 22   & T  & 1.80      & 239   & ind.     \\ \\

J1640+2224   & $-$0.1(3)    & $-$13.3(7)    & C  & 13.3(7)     & 19   & T  & 1.16      & 73    & 2     \\
J1643$-$1224   & 3.0(10)      & $-$8(5)       & C  & 9(5)        & 12   & T  & 2.41      & 103   & 105   \\
B1642$-$03     & $-$3.7(15)     & 30.0(16)    & C  & 30.2(16)    & 1  & I  & 1.12      & 160   & 256   \\
B1700$-$18     & $-$9(11)       & ---         & E  & ---         & 3   & T  & 1.48      & ---      & ---      \\
B1700$-$32     & $-$31(14)      & ---         & E  & ---         & 3   & T  & 2.33      & ---      & ---      \\ \\

B1702$-$19     & $-$66(5)       & ---         & E  & ---         & 3   & T  & 0.88      & ---      & ---      \\
J1709+2313   & $-$3.2(7)      & $-$9.7(9)     & C  & 10.2(9)     & 23   & T  & 1.41      & 68    & 20    \\
B1706$-$16     & 3(9)         & 0(14)    & C  & 3(9)        & 10   & I  & 0.83      & 12    & 6     \\
J1713+0747   & 4.9(3)       & $-$4.1(10)    & C  & 6.4(7)      & 24    & T  & 1.11      & 34    & ind.   \\
B1718$-$02     & $-$1(4)     & $-$26(5)      & C  & 26(5)       & 1  & I  & 2.51      & 309   & 357   \\ \\

B1718$-$32     & 0(5)      & ---         & E  & ---         & 3   & T  & 2.35      & ---      & ---      \\
J1730$-$2304   & 19.9(6)      & ---         & E  & ---         & 3   & T  & 0.53      & ---      & ---      \\
B1732$-$07     & $-$2.4(17)     & 28(3)       & C  & 29(3)       & 1  & I  & 2.26      & 311   & 283   \\
B1735$-$32     & $-$3(13)       & ---         & E  & ---         & 3   & T  & 1.20      & ---      & ---      \\
B1737+13     & $-$22(3)       & $-$20(3)      & C  & 29(3)       & 1  & I  & 1.48      & 203   & 452   \\ \\

J1744$-$1134   & 18.72(6)     & $-$9.5(4)     & C  & 20.99(19)   & 25   & T  & 0.36      & 36    & ind.     \\
B1742$-$30     & 11(4)        & ---         & E  & ---         & 3   & T  & 1.91      & ---      & ---      \\
B1745$-$12     & 3(6)         & ---         & E  & ---         & 3   & T  & 2.45      & ---      & ---      \\
B1749$-$28     & $-$4(6)        & $-$5(5)       & C  & 6(6)        & 10   & I  & 1.23      & 35  & 9     \\
B1754$-$24     & $-$17(5)       & ---         & E  & ---         & 3   & T  & 4.40      & ---      & ---      \\ \\

B1756$-$22     & $-$22(5)       & ---         & E  & ---         & 3   & T  & 3.57      & ---      & ---      \\
B1757$-$24     & 2(9)         & $-$3(9)       & C  & 4(9)        & 26    & I  & 5.22      & 99    & $-$12   \\
B1802$-$07     & $-$5(6)        & ---         & E  & ---         & 3   & T  & 3.10      & ---      & ---      \\
J1804$-$2717   & 2(4)         & ---         & E  & ---         & 3   & T  & 0.78      & ---      & ---      \\
B1802+03     & $-$12(6)       & ---         & E  & ---         & 3   & T  & 2.84      & ---      & ---      \\ \\

B1804$-$08     & $-$5(4)        & 1(4)     & C  & 5(4)        & 10   & I  & 2.73      & 65    & 21    \\
B1806$-$21     & $-$5(7)        & ---         & E  & ---         & 3   & T  & 5.23      & ---      & ---      \\
B1813$-$17     & $-$9(11)       & ---         & E  & ---         & 3   & T  & 6.33      & ---      & ---      \\
B1815$-$14     & $-$12(9)       & ---         & E  & ---         & 3   & T  & 7.15      & ---      & ---      \\
B1818$-$04     & $-$10(3)       & 10(9)       & E  & 14(7)       & 3   & T  & 1.94      & 129   & 34    \\ \\

B1821+05     & 5(11)        & $-$2(4)       & C  & 5(11)       & 4    & I  & 1.83      & 43    & 28    \\
B1820$-$30A    & 2(2)         & ---         & E  & ---         & 3   & T  & 8.00      & ---      & ---      \\
B1820$-$31     & 16(3)        & ---         & E  & ---         & 3   & T  & 1.29      & ---      & ---      \\
B1821$-$24     & $-$0.90(10)    & $-$4.6(18)    & C  & 4.7(18)     & 31    & T  & 5.80      & 129 & ind.     \\
B1822+00     & $-$18(10)      & ---         & E  & ---         & 3   & T  & 1.91      & ---      & ---      \\ \\

B1822$-$09     & $-$13(11)      & $-$9(5)       & C  & 16(10)      & 10   & I  & 0.88      & 67  & 9     \\
B1822$-$14     & 12(14)       & ---         & E  & ---         & 3   & T  & 5.06      & ---      & ---      \\
B1829$-$08     & $-$3(4)        & ---         & E  & ---         & 3   & T  & 4.85      & ---      & ---      \\
B1829$-$10     & $-$13(6)       & ---         & E  & ---         & 3   & T  & 5.81      & ---      & ---      \\
B1830$-$08     & $-$30(3)       & 15(11)      & E  & 34(6)       & 3   & T  & 4.66      & 751   & 163   \\ \\

B1831$-$04     & $-$1(8)     & ---         & E  & ---         & 3   & T  & 2.16      & ---      & ---      \\
B1834$-$04     & $-$2(5)     & 12(14)      & E  & 12(14)      & 3   & T  & 4.94      & 281  & $-$18   \\
B1839+56     & $-$30(4)       & $-$21(2)      & C  & 37(4)       & 4    & I  & 1.68      & 295   & 4     \\
B1838$-$04     & 6(3)         & $-$10(8)      & E  & 12(8)       & 3   & T  & 5.68      & 323   & $-$29  \\
B1842+14     & $-$9(10)       & 45(6)       & C  & 46(7)       & 4    & I  & 2.19      & 478   & 9     \\

\hline \end{tabular}\end{table*}\begin{table*}
\begin{tabular}{llllllllrr}\hline
PSR & $\mu_{\rm long}$ & $\mu_{\rm lat}$ & Coord. & $\mu_{\rm tot}$ & Ref. & Obs. & Dist. & $V_T^{CL}$ & $\Delta V_T$ \\
 & (mas yr$^{-1}$) & (mas yr$^{-1}$) & & (mas yr$^{-1}$) & & Type & (kpc) & (km\,s$^{-1}$) & (km\,s$^{-1}$)\\  
\hline
B1841$-$04     & $-$6(10)       & ---         & E  & ---         & 3   & T  & 2.92      & ---      & ---      \\
B1841$-$05     & 7(5)         & ---         & E  & ---         & 3   & T  & 6.51      & ---      & ---      \\
B1844$-$04     & $-$3(6)        & ---         & E  & ---         & 3   & T  & 3.26      & ---      & ---      \\
B1845$-$01     & $-$1(6)     & ---         & E  & ---         & 3   & T  & 3.99      & ---      & ---      \\
B1846$-$06     & 17(8)        & ---         & E  & ---         & 3   & T  & 3.41      & ---      & ---      \\ \\

B1848+13     & $-$16(6)       & $-$2(10)      & E  & 16(7)       & 3   & T  & 2.69      & 204   & 34    \\
J1852$-$2610   & $-$15(11)      & ---         & E  & ---         & 3   & T  & 1.75      & ---      & ---      \\
B1855+02     & $-$11(7)       & ---         & E  & ---         & 3   & T  & 7.98      & ---      & ---      \\
B1855+09     & $-$2.94(4)     & $-$5.41(6)    & C  & 6.16(6)     & 27    & T  & 0.91      & 27    & ind.     \\
B1857$-$26     & $-$19.9(3)     & $-$47.3(9)    & C  & 51.3(9)     & 28   & I  & 2.00      & 486   & ind.     \\ \\

B1859+01     & 6(10)        & ---         & E  & ---         & 3   & T  & 2.79      & ---      & ---      \\
B1900+05     & $-$8(8)        & ---         & E  & ---         & 3   & T  & 4.66      & ---      & ---      \\
B1900+06     & $-$4(6)        & $-$7(13)      & E  & 8(12)       & 3   & T  & 8.44      & 320   & 45    \\
B1900$-$06     & $-$2(8)        & ---         & E  & ---         & 3   & T  & 5.37      & ---      & ---      \\
B1902$-$01     & 1(5)      & 9(14)       & E  & 9(14)       & 3   & T  & 6.01      & 256   & 38    \\ \\

B1904+06     & $-$6(6)        & $-$1(11)   & E  & 6(7)        & 3   & T  & 8.31      & 236   & 26    \\
B1905+39     & 11(4)        & 11.0(10)    & C  & 16(3)       & 4    & I  & 2.24      & 170   & $-$36   \\
B1907+00     & $-$2(6)     & ---         & E  & ---         & 3   & T  & 3.57      & ---      & ---      \\
B1907+02     & 1(10)     & ---         & E  & ---         & 3   & T  & 4.94      & ---      & ---      \\
B1907+10     & $-$5(4)        & 8(8)        & E  & 9(8)        & 3   & T  & 4.18      & 178   & 6     \\ \\

J1909$-$3744   & $-$9.6(2)      & $-$35.6(7)    & C  & 36.9(7)     & 29   & T  & 0.46      & 80    & 86    \\
J1911$-$1114   & $-$6(4)        & $-$23(13)     & C  & 24(13)      & 12   & T  & 1.22      & 139   & 42    \\
B1911+13     & $-$8(5)        & $-$11(8)      & E  & 14(8)       & 3   & T  & 5.12      & 340   & $-$3    \\
B1911$-$04     & 7(13)        & $-$5(9)       & C  & 9(12)       & 4    & I  & 2.79      & 119   & 18    \\
B1913+10     & 1(3)      & $-$7(6)       & E  & 7(6)        & 3   & T  & 6.27      & 208   & $-$32   \\ \\

B1913+16     & $-$2.56(6)     & 0.49(7)     & C  & 2.61(7)     & 30     & T  & 5.90      & 73    & 15    \\
B1914+09     & $-$9(4)        & $-$1(7)    & E  & 9(5)        & 3   & T  & 2.94      & 125   & $-$3    \\
B1917+00     & ---          & $-$1(10)   & C  & ---         & 4    & I  & 3.06      & ---      & ---      \\
B1919+21     & ---          & 40(10)      & C  & ---         & 10   & I  & 1.09      & ---      & ---      \\
B1924+16     & 3(8)         & $-$16(13)     & E  & 16(13)      & 3   & T  & 5.83      & 442   & 145   \\ \\

B1929+10     & 94.03(14)    & 43.4(3)     & C  & 103.56(18)  & 8   & I  & 0.36      & 177   & ind.   \\
B1933+16     & $-$1(3)     & $-$13(3)      & C  & 13(3)       & 10   & I  & 5.61      & 346   & 143   \\
B1935+25     & $-$14(3)       & $-$11(4)      & E  & 18(4)       & 3   & T  & 3.25      & 277   & $-$42   \\
B1937+21     & $-$0.46(2)     & $-$0.66(2)    & C  & 0.80(2)     & 32   & T  & 3.60      & 14    & ind.     \\
B1937$-$26     & 12(2)        & $-$10(4)      & C  & 16(3)       & 33   & I  & 1.66      & 126   & 234   \\ \\

B1944+17     & 1(5)      & $-$9(4)       & C  & 9(5)        & 6    & I  & 1.40      & 60    & $-$23   \\
B1943$-$29     & 19(9)        & ---         & C  & ---         & 1  & I  & 1.54      & ---      & ---      \\
B1946+35     & $-$12.6(6)     & $-$0.7(6)     & C  & 12.6(6)     & 4?   & I  & 5.80      & 346   & 124   \\
B1951+32     & $-$24(4)       & $-$8(3)       & C  & 25(4)       & 34   & I  & 2.50      & 296   & ind.     \\
B1952+29     & ---          & $-$36(10)     & C  & ---         & 6    & I  & 0.70      & ---      & ---      \\ \\

B1953+29     & $-$1.0(3)      & $-$3.7(3)     & C  & 3.8(3)      & 19   & T  & 4.64      & 84    & 14    \\
B1953+50     & $-$23(5)       & 54(5)       & C  & 59(5)       & 4    & I  & 2.24      & 626   & $-$123  \\
B1957+20     & $-$16.0(5)     & $-$25.8(6)    & C  & 30.4(6)     & 35    & T  & 2.49      & 359   & $-$138  \\
B2000+40     & $-$13(7)       & $-$3(7)       & E  & 13(7)       & 3   & T  & 5.90      & 364   & 148   \\
B2002+31     & $-$5(8)        & $-$10(11)     & E  & 11(11)      & 3   & T  & 7.50      & 391  & 75    \\ \\

B2011+38     & $-$32.1(17)    & $-$25(3)      & C  & 41(2)    & 1  & I  & 8.44      & 1624  & 891   \\
B2016+28     & $-$2.6(2)      & $-$6.2(4)     & C  & 6.7(4)      & 7   & I  & 0.97      & 31    & ind.     \\
J2019+2425   & $-$9.41(12)    & $-$20.60(15)  & C  & 22.65(15)   & 36    & T  & 1.49      & 160  & $-$62   \\
B2020+28     & $-$4.4(5)      & $-$23.6(3)    & C  & 24.0(4)     & 7   & I  & 2.70      & 307   & ind.     \\
B2021+51     & $-$5.23(17)    & 11.5(3)     & C  & 12.6(3)     & 7   & I  & 2.00      & 119   & ind.     \\

\hline \end{tabular}\end{table*}\begin{table*}
\begin{tabular}{llllllllrr}\hline
PSR & $\mu_{\rm long}$ & $\mu_{\rm lat}$ & Coord. & $\mu_{\rm tot}$ & Ref. & Obs. & Dist. & $V_T^{CL}$ & $\Delta V_T$ \\
 & (mas yr$^{-1}$) & (mas yr$^{-1}$) & & (mas yr$^{-1}$) & & Type & (kpc) & (km\,s$^{-1}$) & (km\,s$^{-1}$)\\  
\hline
B2022+50     & 17(3)        & 14(3)       & E  & 22(3)       & 3   & T  & 2.34      & 244   & $-$56   \\
B2044+15     & $-$13(6)       & 3(4)        & C  & 13(6)       & 4    & I  & 2.42      & 149   & 9     \\
B2043$-$04     & ---          & $-$7(8)       & C  & ---         & 4    & I  & 1.75      & ---      & ---      \\
B2045$-$16     & 117(5)       & $-$5(5)       & C  & 117(5)      & 10   & I  & 0.56      & 311   & 44    \\
J2051$-$0827   & 5.3(10)      & 0(3)     & C  & 5.3(11)     & 37   & T  & 1.04      & 26    & 6     \\ \\

B2053+21     & $-$5(6)        & $-$1(9)    & E  & 5(7)        & 3   & T  & 2.43      & 58   & $-$7    \\
B2053+36     & $-$3(7)        & 3(3)        & C  & 4(6)        & 4    & I  & 4.62      & 88    & 18    \\
B2106+44     & 3.5(13)      & 1.4(14)     & C  & 3.8(14)     & 1  & I  & 4.96      & 89   & 6     \\
B2110+27     & $-$23(2)       & $-$54(3)      & C  & 59(3)       & 4    & I  & 2.03      & 568   & $-$179  \\
B2111+46     & 7(12)        & $-$11(13)     & E  & 13(13)      & 3   & T  & 4.53      & 279   & 28    \\ \\

B2113+14     & ---          & $-$11(5)      & C  & ---         & 4    & I  & 4.17      & ---      & ---      \\
J2124$-$3358   & $-$14.0(10)    & $-$47.0(10)   & C  & 49.0(10)    & 12   & T  & 0.27      & 63    & $-$5   \\
J2129$-$5721   & 7(2)         & $-$4(3)       & C  & 8(3)        & 12   & T  & 1.36      & 52    & 45    \\
J2145$-$0750   & $-$12.37(6)      & $-$6.8(8)  & C  & 14.1(4)    & 40   & T  & 0.50      & 33      & ind.      \\
B2148+63     & 14(3)        & 10(4)       & C  & 17(4)       & 4    & I  & 5.51      & 444   & 656   \\ \\

B2148+52     & 7(3)         & $-$5(3)       & E  & 9(3)        & 3   & T  & 4.62      & 197   & 45    \\
B2154+40     & 17.8(8)      & 2.8(10)     & C  & 18.0(9)     & 1  & I  & 3.76      & 321   & 155   \\
B2217+47     & $-$12(8)       & $-$30(6)      & C  & 32(7)       & 6    & I  & 2.22      & 337   & 35    \\
B2224+65     & 144(3)       & 112(3)      & C  & 182(3)      & 4    & I  & 1.86      & 1605  & 121   \\
J2229+2643   & 1(4)      & $-$17(4)      & C  & 17(4)       & 19   & T  & 1.45      & 117   & $-$2    \\ \\

B2227+61     & 8(10)        & 6(11)       & E  & 10(11)      & 3   & T  & 4.04      & 192   & 79    \\
J2235+1506   & 15(4)        & 10(8)       & C  & 18(6)       & 38    & T  & 1.07      & 91    & 7     \\
B2255+58     & 11(4)        & $-$14(5)      & E  & 18(5)       & 3   & T  & 4.51      & 385   & 161   \\
B2303+30     & 2(2)         & $-$20(2)      & C  & 20(3)       & 1  & I  & 3.66      & 347   & 25    \\
B2306+55     & $-$15(8)       & ---         & C  & ---         & 4    & I  & 2.16      & ---      & ---      \\ \\

B2310+42     & 21.4(9)      & $-$5.3(12)    & E  & 22.0(10)    & 3   & T  & 1.25      & 130   & $-$31   \\
J2317+1439   & $-$1.7(15)     & 7(4)        & C  & 8(4)        & 38    & T  & 0.83      & 31    & 40    \\
B2315+21     & 2(9)      & ---         & E  & ---         & 3   & T  & 0.95      & ---      & ---      \\
J2322+2057   & $-$17(2)       & $-$18(3)      & C  & 25(3)       & 39     & T  & 0.80      & 95    & $-$2    \\
B2324+60     & $-$18(5)       & 6(5)        & E  & 19(5)       & 3   & T  & 4.86      & 438   & $-$4   \\ \\

B2327$-$20     & 74.7(19)     & 5(3)        & C  & 75(2)    & 1  & I  & 0.39      & 138   & 36    \\
B2334+61     & $-$7(12)       & $-$8(13)      & E  & 11(13)      & 3   & T  & 3.15      & 164   & $-$35  \\
B2351+61     & 22(3)        & 6(2)        & C  & 23(3)       & 4    & I  & 3.43      & 374   & $-$13   \\
\hline \end{tabular}

References: 1. Brisken et al. (2003a)\nocite{bfg+03a}, 2. Freire et
al. (2003)\nocite{fck+03}, 3. Hobbs et al. (2004)\nocite{hlk+04},
4. Harrison, Lyne \& Anderson (1993)\nocite{hla93}, 5. Siegman,
Manchester \& Durdin (1993)\nocite{smd93}, 6. Lyne, Anderson \& Salter
(1982)\nocite{las82}, 7. Brisken et al. (2002)\nocite{bbgt02},
8. Chatterjee et al. (2004b)\nocite{ccv+04}, 
9. van Straten et al. (2001)\nocite{vbb+01}, 
10. Fomalont et al. (1997)\nocite{fgml97}, 
11. Wyckoff \& Murray (1977)\nocite{wm77}
12. Toscano et al. (1999b)\nocite{tsb+99},
13. Splaver et al. (2002)\nocite{sna+02}, 
14. Caraveo et al. (1996)\nocite{cbmt96}, 
15. Brisken et al. (2003b)\nocite{btgg03b}
16. Dodson et al. (2003)\nocite{dlrm03}, 
17. Chatterjee et al. (2001)\nocite{ccl+01}, 
18. Lange et al. (2001)\nocite{lcw+01},
19. Wolszczan et al. (2000)\nocite{wdk+00}, 
20. Bailes et al. (1990a)\nocite{bmk+90b},
21. Konacki, Wolszczan \& Stairs (2003)\nocite{kws03},
22. Thorsett et al. (1999)\nocite{tacl99},
23. Lewandowski et al. (2004)\nocite{lwf+04},
24. Camilo, Foster \& Wolszczan (1994)\nocite{cfw94},
25. Toscano et al. (1999a)\nocite{tbm+99},
26. Thorsett, Birsken \& Goss (2002)\nocite{tbg02},
27. Kaspi, Taylor \& Ryba (1994)\nocite{ktr94},
28. Romalont et al. (1999)\nocite{fgbc99},
29. Jacoby et al. (2003)\nocite{jbv+03},
30. Weisberg \& Taylor (2003)\nocite{wt03}
31. Cognard \& Kestrade (1997)\nocite{cl97a},
32. Cognard et al. (1995)\nocite{cbl+95},
33. McGary et al. (2001)\nocite{mbf+01},
34. Migliazzo et al. (2002)\nocite{mgb+02},
35. Arzoumanian, Fruchter \& Taylor (1994)\nocite{aft94},
36. Nice, Splaver \& Stairs (2001)\nocite{nss01},
37. Doroshenko et al. (2001)\nocite{dlk+01},
38. Camilo, Nice \& Taylor (1996)\nocite{cnt96},
39. Nice \& Taylor (1995)\nocite{nt95},
40. L{\"o}hmer et al. (2004)\nocite{lkd+04}.
\end{table*}

 The resulting sample of 233 proper motions is provided in
 Table~\ref{tb:pmsample}. As detailed in the caption,
 Table~\ref{tb:pmsample} also includes the 2-D speed (often
 referred to as the transverse speed), $V_T^{CL}
 = \mu_{tot}D^{CL}$, and the difference in 2-D speed between
 CL02 and TC93, $V_T^{TC}-V_T^{CL}$.  

 The sample contains a wide variety of pulsars including `normal',
 recycled (defined here as pulsars with spin-periods, $P < 0.1$\,s and
 spin-down rates, $\dot{P} < 10^{-17}$), `young' (those pulsars with
 characteristic ages $\tau_c < 3$\,Myr) and pulsars associated with
 supernova remnants or globular clusters.  In our sample, 58\% of the
 proper motions were determined using timing methods, 41\% using
 interferometers and the remaining 1\% using optical telescopes.  In
 Figure~\ref{fg:aitoff} we plot the positions and proper motions on
 the sky. The apparent motions with respect to the Galactic plane seen
 in this figure are discussed in detail in Section \ref{sec:gplane}.

\begin{table*}\caption{Proper motions in Galactic coordinates.  In
column order, the table lists pulsar name, Galactic coordinates,
proper motion in Galactic coordinates, the height above the Galactic
plane, the velocity in this direction, $V_Z = \mu_b D \cos b$, the base-10 logarithms of the 
characteristic age and surface magnetic field
strength.}\label{tb:gmotion}\begin{tabular}{lrrllrrll}\hline
PSR & $l$ & $b$& $\mu_l$ & $\mu_b$ & Z & $V_Z$ & log[$\tau_c$ (yr)]& log[$B_s$ (G)] \\
 & ($^\circ$) & ($^\circ$) & (mas yr$^{-1}$) & (mas yr$^{-1}$) & (kpc) & (km s$^{-1}$) &  &  \\
\hline
B0011+47     & 116.57  & $-$14.64  & 18.4(18)        & $-$20.8(16)    & $-$0.39  & $-$149  & 7.54   & 11.93   \\
B0021$-$72C    & 305.99  & $-$44.88  & $-$1.4(5)         & 3.6(5)       & $-$3.39  & 58    & ---   & ---    \\
B0021$-$72D    & 305.94  & $-$44.88  & 0.1(4)        & 3.1(4)       &    $-$3.39  & 50    & ---   & ---    \\
B0021$-$72E    & 305.95  & $-$44.87  & $-$2.4(5)         & 3.0(5)       & $-$3.39  & 48    & 8.76   & 8.78    \\
B0021$-$72F    & 305.96  & $-$44.88  & $-$0.8(6)         & 3.4(6)       & $-$3.39  & 55    & 8.81   & 8.62    \\ \\

B0021$-$72G    & 305.95  & $-$44.88  & $-$0.5(14)        & 4.3(13)      & $-$3.39  & 69    & ---   & ---    \\
B0021$-$72H    & 305.96  & $-$44.89  & $-$1.4(12)        & 4.1(12)      & $-$3.39  & 66    & ---   & ---    \\
B0021$-$72I    & 305.95  & $-$44.88  & $-$1.2(17)        & 4.4(19)      & $-$3.39  & 71    & ---   & ---    \\
B0021$-$72J    & 305.97  & $-$44.89  & $-$1.62(14)       & 3.84(15)     & $-$3.39  & 62    & ---   & ---    \\
B0021$-$72N    & 305.95  & $-$44.88  & $-$2.7(14)        & 1.3(14)      & $-$3.39  & 21    & ---   & ---    \\ \\

J0024$-$7204O  & 305.96  & $-$44.88  & $-$0.8(11)        & 3.4(10)      & $-$3.39  & 55    & 9.14   & 8.46    \\
J0024$-$7204U  & 305.95  & $-$44.89  & $-$1.5(7)         & 5.0(7)       & $-$3.39  & 81    & 8.86   & 8.81    \\
J0034$-$0534   & 111.58  & $-$68.08  & $-$27(9)          & 15(3)        & $-$0.50  & 14    & 9.78   & 7.99    \\
B0037+56     & 121.53  & $-$5.58   & $-$6(12)          & $-$2(11)    & $-$0.30  & $-$22   & 6.79   & 12.26   \\
B0114+58     & 126.36  & $-$3.47   & 10(6)           & 17(6)        & $-$0.13  & 179   & 5.44   & 11.89   \\ \\

J0134$-$2937   & 230.38  & $-$80.24  & $-$8(5)           & 20(4)        & $-$0.55  & 9     & 7.44   & 11.02   \\
B0136+57     & 129.29  & $-$4.06   & $-$6(5)           & $-$20(5)       & $-$0.20  & $-$272  & 5.61   & 12.24   \\
B0144+59     & 130.13  & $-$2.73   & 12(3)           & 4(3)         & $-$0.11  & 42    & 7.08   & 11.36   \\
B0149$-$16     & 179.42  & $-$72.46  & 20.2(18)        & $-$16.6(16)    & $-$0.49  & $-$12   & 7.01   & 12.02   \\
J0218+4232   & 139.58  & $-$17.54  & $-$2(7)           & 2(4)      & $-$0.80  & 22    & 8.68   & 8.63    \\ \\

B0301+19     & 161.22  & $-$33.27  & 24(7)           & $-$25(6)       & $-$0.34  & $-$61   & 7.23   & 12.13   \\
B0320+39     & 152.26  & $-$14.34  & 29(6)           & $-$14(6)       & $-$0.25  & $-$65   & 7.88   & 12.15   \\
B0329+54     & 145.07  & $-$1.23   & 18.2(4)         & 3.5(4)       & $-$0.02  & 18    & 6.74   & 12.09   \\
B0331+45     & 150.42  & $-$8.04   & 9(8)            & $-$1(4)     & $-$0.23  & $-$10   & 8.76   & 10.65   \\
B0353+52     & 149.17  & $-$0.53   & $-$8(6)           & 1(4)      & $-$0.03  & 13    & 6.82   & 11.49   \\ \\

B0355+54     & 148.26  & 0.80    & 0.4(3)          & 13.5(4)      & 0.02   & 70    & 5.75   & 11.92   \\
B0402+61     & 144.10  & 7.04    & 40(11)          & 54(9)        & 0.26   & 539   & 6.23   & 12.26   \\
B0410+69     & 138.98  & 13.66   & $-$3(7)           & $-$11(7)       & 0.28   & $-$60   & 7.91   & 11.24   \\
J0437$-$4715   & 253.47  & $-$41.95  & 75.456(9)       & 120.263(9)   & $-$0.09  & 59    & 9.20   & 8.76    \\
B0450+55     & 152.69  & 7.54    & 43(5)           & 32(5)        & 0.09   & 101   & 6.36   & 11.96   \\ \\

B0458+46     & 160.43  & 3.07    & $-$12(5)          & $-$1(4)     & 0.07   & $-$5    & 6.26   & 12.28   \\
B0523+11     & 192.77  & $-$13.24  & 19(6)           & 23(7)        & $-$0.71  & 330   & 7.88   & 11.21   \\
B0531+21     & 184.63  & $-$5.78   & $-$14(3)          & $-$7(3)        & $-$0.20  & $-$66   & 3.09   & 12.58   \\
B0540+23     & 184.44  & $-$3.32   & $-$1(8)        & 23(8)        & $-$0.12  & 224   & 5.40   & 12.29   \\
B0559$-$05     & 212.27  & $-$13.47  & 23(8)           & 9(8)         & $-$0.92  & 163   & 6.68   & 11.86   \\ \\

B0609+37     & 175.52  & 9.09    & 7(12)           & 6(5)         & 0.14   & 24    & 7.90   & 11.13   \\
J0613$-$0200   & 210.49  & $-$9.30   & 7.8(10)         & $-$1.3(6)      & $-$0.28  & $-$10   & 9.71   & 8.24    \\
B0611+22     & 188.86  & 2.40    & 0(7)         & $-$5(6)        & 0.09   & $-$49   & 4.95   & 12.66   \\
J0621+1002   & 200.64  & $-$2.01   & 1.7(9)          & 3.6(5)       & $-$0.05  & 23    & 9.98   & 9.07    \\
B0626+24     & 188.89  & 6.23    & $-$5(12)          & $-$5(12)       & 0.24   & $-$53   & 6.58   & 11.99   \\ \\

B0628$-$28     & 237.03  & $-$16.75  & $-$32(3)          & $-$34.3(12)    & $-$0.42  & $-$226  & 6.44   & 12.48   \\
J0633+1746   & 195.21  & 4.27    & $-$33(4)          & 178(4)       & 0.01   & 135   & 5.53   & 12.21   \\
B0643+80     & 133.24  & 26.82   & 3(3)            & 18(3)        & 0.70   & 118   & 6.71   & 12.34   \\
B0656+14     & 201.18  & 8.26    & 17(3)           & 44(3)        & 0.04   & 60    & 5.05   & 12.67   \\
B0655+64     & 151.62  & 25.23   & $-$1(7)        & $-$10(7)       & 0.21   & $-$21   & 9.66   & 10.07   \\ \\

J0711$-$6830   & 279.60  & $-$23.27  & $-$11.6(6)        & $-$10.3(6)     & $-$0.34  & $-$39   & 9.77   & 8.46    \\
B0736$-$40     & 254.27  & $-$9.18   & 7(2)         & $-$18.6(15)    & $-$0.42  & $-$230  & 6.57   & 11.90   \\
B0740$-$28     & 243.85  & $-$2.43   & $-$13.6(2)        & $-$22.6(2)     & $-$0.09  & $-$222  & 5.20   & 12.23   \\
B0751+32     & 188.25  & 26.72   & $-$8(4)           & $-$1(5)     & 0.68   & $-$6    & 7.33   & 12.10   \\
B0756$-$15     & 234.54  & 7.24    & 0(5)         & 4(5)         & 0.37   & 55    & 6.82   & 12.03   \\

\hline \end{tabular}\end{table*}\begin{table*}
\begin{tabular}{lrrllrrll}\hline
PSR & $l$ & $b$& $\mu_l$ & $\mu_b$ & Z & $V_Z$ & log[$\tau_c$ (yr)]& log[$B_s$ (G)] \\
 & ($^\circ$) & ($^\circ$) & (mas yr$^{-1}$) & (mas yr$^{-1}$) & (kpc) & (km s$^{-1}$) &  &  \\
\hline
B0809+74     & 140.06  & 31.61   & 32.8(4)         & 34.69(13)    & 0.23   & 60    & 8.09   & 11.67   \\
B0818$-$13     & 235.96  & 12.61   & 48(7)           & $-$16(9)       & 0.43   & $-$147  & 6.97   & 12.21   \\
B0820+02     & 222.06  & 21.26   & 5(9)            & 6(11)        & 0.37   & 27    & 8.12   & 11.48   \\
B0823+26     & 197.03  & 31.75   & 100(3)          & 40(3)        & 0.19   & 58    & 6.69   & 11.98   \\
B0833$-$45     & 263.62  & $-$2.77   & $-$42.21(9)       & $-$17.25(8)    & $-$0.01  & $-$24   & 4.05   & 12.53   \\ \\

B0834+06     & 219.79  & 26.28   & $-$43(4)          & 30(5)        & 0.29   & 83    & 6.47   & 12.47   \\
B0835$-$41     & 260.98  & $-$0.32   & 18(3)           & $-$12(3)       & $-$0.01  & $-$91   & 6.53   & 12.22   \\
B0906$-$17     & 246.19  & 19.86   & 54(11)          & 1(11)     & 0.31   & 2     & 6.98   & 11.72   \\
B0919+06     & 225.48  & 36.40   & $-$64.95(11)      & 61.69(8)     & 0.71   & 282   & 5.70   & 12.39   \\
B0950+08     & 228.97  & 43.71   & $-$24.23(8)       & 26.20(8)     & 0.18   & 23    & 7.24   & 11.39   \\ \\

J1012+5307   & 160.40  & 50.85   & 16.8(2)         & 15.6(2)      & 0.32   & 19    & 9.69   & 8.48    \\
J1024$-$0719   & 251.77  & 40.53   & 27(3)           & $-$70(3)       & 0.25   & $-$98   & 9.64   & 8.50    \\
B1039$-$19     & 265.66  & 33.60   & $-$4(4)           & 12(4)        & 0.82   & 70    & 7.37   & 12.06   \\
J1045$-$4509   & 280.92  & 12.27   & $-$1.1(19)        & 3.5(13)      & 0.42   & 32   & 9.83   & 8.57    \\
B1112+50     & 154.46  & 60.36   & 17(3)           & 53(3)        & 0.39   & 56    & 7.02   & 12.31   \\ \\

B1133+16     & 241.96  & 69.21   & $-$338.0(4)       & 166.0(4)     & 0.34   & 101   & 6.70   & 12.33   \\
B1237+25     & 252.45  & 86.56   & $-$105.92(18)     & $-$44.91(18)   & 0.86   & $-$11   & 7.36   & 12.07   \\
B1257+12     & 311.41  & 75.42   & 39.68(11)       & $-$87.4(2)   & 0.75   & $-$80   & 8.94   & 8.93    \\
B1322+83     & 121.95  & 33.66   & 54(20)          & $-$6(8)        & 0.42   & $-$18   & 7.27   & 11.79   \\
B1325$-$43     & 309.95  & 18.43   & 17(8)           & 53(23)       & 0.42   & 319   & 6.45   & 12.11   \\ \\

B1426$-$66     & 312.72  & $-$5.39   & $-$31(5)          & $-$7(4)        & $-$0.09  & $-$33   & 6.65   & 12.17   \\
B1449$-$64     & 315.80  & $-$4.42   & $-$18.9(10)       & $-$10.7(9)     & $-$0.16  & $-$105  & 6.02   & 11.85   \\
J1455$-$3330   & 330.80  & 22.57   & 22(8)           & 19(11)       & 0.20   & 44    & 9.72   & 8.65    \\
B1451$-$68     & 313.94  & $-$8.53   & $-$31.2(4)        & 11.0(4)      & $-$0.07  & 23    & 7.63   & 11.21   \\
B1508+55     & 91.40   & 52.27   & $-$16.9(15)       & 96.0(14)     & 0.78   & 276   & 6.37   & 12.29   \\ \\

J1518+4904   & 80.88   & 54.27   & 6.5(6)          & $-$7.8(6)      & 0.51   & $-$14  & 10.37  & 9.03    \\
B1534+12     & 19.94   & 48.34   & $-$19.517(19)     & $-$12.922(13)  & 0.71   & $-$39   & 8.39   & 9.99    \\
B1541+09     & 17.90   & 45.77   & $-$7.2(10)        & 0.6(10)      & 2.50   & 7     & 7.44   & 11.76   \\
B1540$-$06     & 0.65    & 36.61   & $-$12(3)          & 11(3)        & 0.43   & 30    & 7.11   & 11.90   \\
B1556$-$44     & 334.61  & 6.37    & 13(9)           & 10(10)       & 0.25   & 108   & 6.60   & 11.71   \\ \\

J1603$-$7202   & 316.70  & $-$14.49  & $-$2.9(5)         & $-$1.7(5)      & $-$0.29  & $-$9    & 10.18  & 8.69    \\
B1600$-$49     & 332.22  & 2.45    & $-$20(6)          & 19(6)        & 0.22   & 459   & 6.71   & 11.77   \\
B1604$-$00     & 10.80   & 35.46   & $-$3(11)          & $-$3(13)       & 0.39   & $-$8    & 7.34   & 11.56   \\
B1620$-$26     & 351.05  & 15.96   & $-$26(4)          & $-$7(4)        & 0.49   & $-$57   & 8.42   & 9.44    \\
J1640+2224   & 41.13   & 38.26   & $-$10.2(7)        & $-$3.1(4)      & 0.72   & $-$13   & 10.25  & 7.98    \\ \\

J1643$-$1224   & 5.75    & 21.22   & $-$4(5)           & $-$7(4)        & 0.87   & $-$75   & 9.60   & 8.47    \\
B1642$-$03     & 14.19   & 26.06   & 24.4(16)        & 20.0(16)     & 0.49   & 95    & 6.54   & 11.92   \\
J1709+2313   & 44.60   & 32.20   & $-$7.3(9)         & 1.1(8)       & 0.75   & 6     & 10.31  & 8.12    \\
B1706$-$16     & 5.85    & 13.66   & 3(13)           & $-$2(11)    & 0.20   & $-$6    & 6.21   & 12.31   \\
J1713+0747   & 28.83   & 25.22   & 0.8(10)         & $-$5.1(6)      & 0.47   & $-$24   & 9.93   & 8.30    \\ \\

B1718$-$02     & 20.21   & 18.93   & $-$20(5)          & $-$11(4)       & 0.81   & $-$124  & 7.96   & 11.30   \\
B1732$-$07     & 17.35   & 13.28   & 25(3)           & 17(2)     & 0.52   & 179   & 6.74   & 11.86   \\
B1737+13     & 37.16   & 21.67   & $-$23(3)          & 13(3)        & 0.55   & 85    & 6.94   & 12.04   \\
J1744$-$1134   & 14.87   & 9.18    & 6.2(4)          & $-$17.8(3)     & 0.06   & $-$30   & 9.86   & 8.29    \\
B1749$-$28     & 1.61    & $-$0.96   & $-$5(6)           & 2(6)      & $-$0.02  & 11   & 6.04   & 12.33   \\ \\

B1757$-$24     & 5.29    & $-$0.90   & $-$1(9)        & $-$3(9)        & $-$0.08  & $-$74   & 4.19   & 12.61   \\
B1804$-$08     & 20.14   & 5.58    & 0(4)         & 5(4)         & 0.27   & 64    & 7.95   & 10.84   \\
B1818$-$04     & 25.53   & 4.73    & $-$1(9)        & 13(5)        & 0.16   & 119   & 6.18   & 12.29   \\
B1821+05     & 35.06   & 8.85    & 4(7)            & $-$5(10)       & 0.28   & $-$43   & 7.72   & 11.62   \\
B1821$-$24     & 7.87    & $-$5.58   & $-$3.6(16)        & $-$1.2(9)      & $-$0.56  & $-$33   & 7.48   & 9.35    \\

\hline \end{tabular}\end{table*}\begin{table*}
\begin{tabular}{lrrllrrll}\hline
PSR & $l$ & $b$& $\mu_l$ & $\mu_b$ & Z & $V_Z$ & log[$\tau_c$ (yr)]& log[$B_s$ (G)] \\
 & ($^\circ$) & ($^\circ$) & (mas yr$^{-1}$) & (mas yr$^{-1}$) & (kpc) & (km s$^{-1}$) &  &  \\
\hline
B1822$-$09     & 21.52   & 1.32    & $-$12(7)          & 9(11)        & 0.02   & 38    & 5.37   & 12.81   \\
B1830$-$08     & 23.46   & 0.06    & $-$2(11)       & 33(5)        & 0.00   & 729   & 5.17   & 11.95   \\
B1834$-$04     & 27.24   & 1.12    & $-$6(14)          & 5(7)         & 0.10   & 117   & 6.53   & 11.89   \\
B1839+56     & 86.15   & 23.80   & $-$23(3)          & 24(4)        & 0.68   & 175   & 7.24   & 12.20   \\
B1838$-$04     & 27.89   & 0.27    & 11(8)           & $-$10(5)       & 0.03   & $-$269  & 5.66   & 12.04   \\ \\

B1842+14     & 45.63   & 8.14    & 40(7)           & 29(10)       & 0.31   & 298   & 6.50   & 11.93   \\
B1848+13     & 45.06   & 6.33    & 17(9)           & 9(9)         & 0.30   & 114   & 6.56   & 11.86   \\
B1855+09     & 42.36   & 3.05    & $-$2.94(6)        & 1.74(5)      & 0.05   & 7     & 9.68   & 8.50    \\
B1857$-$26     & 10.41   & $-$13.45  & $-$50.1(9)        & $-$0.3(5)      & $-$0.47  & $-$3    & 7.68   & 11.55   \\
B1900+06     & 39.89   & 0.33    & 11(12)          & $-$1.8(91)     & 0.05   & $-$72   & 6.14   & 12.36   \\ \\

B1902$-$01     & 33.76   & $-$3.56   & $-$5(13)          & 2(8)         & $-$0.37  & 57    & 6.52   & 12.15   \\
B1904+06     & 40.68   & $-$0.31   & 8(10)           & 3(9)         & $-$0.04  & 118   & 6.30   & 11.88   \\
B1905+39     & 71.02   & 14.19   & 19.1(19)        & $-$5(4)        & 0.55   & $-$51   & 7.56   & 11.92   \\
B1907+10     & 44.91   & 0.98    & 1(7)         & 8(7)         & 0.07   & 158   & 6.23   & 11.94   \\
J1909$-$3744   & 359.80  & $-$19.60  & $-$34.0(7)        & $-$0.5(4)      & $-$0.28  & $-$2    & 9.52   & 8.32    \\ \\

J1911$-$1114   & 25.21   & $-$9.59   & $-$21(12)         & $-$4(7)        & $-$0.20  & $-$23   & 9.61   & 8.36    \\
B1911+13     & 47.95   & 1.58    & 17(7)           & $-$4(7)        & 0.14   & $-$97   & 7.01   & 11.82   \\
B1911$-$04     & 31.38   & $-$7.13   & 2(10)        & $-$8(13)       & $-$0.35  & $-$105  & 6.51   & 12.27   \\
B1913+10     & 44.78   & $-$0.66   & 8(5)            & $-$6(5)        & $-$0.07  & $-$178  & 5.62   & 12.40   \\
B1913+16     & 50.04   & 2.11    & 5.16(7)         & 2.73(7)      & 0.22   & 76    & 8.03   & 10.36   \\ \\

B1914+09     & 44.63   & $-$1.03   & 10(6)           & 5(6)         & $-$0.05  & 70    & 6.23   & 11.92   \\
B1924+16     & 51.93   & 0.05    & 9(10)           & $-$16(12)      & 0.01   & $-$442  & 5.71   & 12.51   \\
B1929+10     & 47.45   & $-$3.90   & 86.5(3)         & $-$57.41(19)   & $-$0.02  & $-$98   & 6.49   & 11.71   \\
B1933+16     & 52.51   & $-$2.10   & $-$6(3)           & $-$5(3)        & $-$0.21  & $-$133  & 5.98   & 12.17   \\
B1935+25     & 60.91   & 2.26    & 20(4)           & $-$7(4)        & 0.13   & $-$108  & 6.69   & 11.56   \\ \\

B1937+21     & 57.58   & $-$0.30   & 4.40(2)         & 0.47(2)      & $-$0.02  & 8     & 8.37   & 8.61    \\
B1937$-$26     & 13.97   & $-$21.83  & $-$4(4)           & $-$15(3)       & $-$0.62  & $-$110  & 6.82   & 11.80   \\
B1944+17     & 55.40   & $-$3.51   & $-$4(5)           & $-$4(5)        & $-$0.09  & $-$26   & 8.46   & 11.02   \\
B1946+35     & 70.77   & 5.03    & $-$1.8(6)         & 10.7(6)      & 0.51   & 293   & 6.21   & 12.36   \\
B1951+32     & 68.84   & 2.81    & $-$14(4)          & 17(4)        & 0.12   & 201   & 5.03   & 11.69   \\ \\

B1953+29     & 65.91   & 0.43    & 1.9(3)          & $-$0.8(3)      & 0.03   & $-$18   & 9.51   & 8.64    \\
B1953+50     & 84.87   & 11.54   & 40(5)           & 47(5)        & 0.45   & 489   & 6.78   & 11.93   \\
B1957+20     & 59.27   & $-$4.71   & $-$26.0(6)        & 1.1(6)       & $-$0.20  & 13    & 9.18   & 8.22    \\
B2000+40     & 76.68   & 5.27    & 13(7)           & $-$5(7)        & 0.54   & $-$139  & 6.92   & 12.10   \\
B2002+31     & 69.08   & 0.01    & 7(9)            & $-$11(11)      & 0.00   & $-$391  & 5.65   & 13.10   \\ \\

B2011+38     & 76.00   & 2.46    & $-$35(3)          & 13(2)     & 0.36   & 523.59   & 5.61   & 12.16   \\
B2016+28     & 68.17   & $-$4.00   & $-$2.5(4)         & 0.3(3)       & $-$0.07  & 1.38     & 7.78   & 11.46   \\
J2019+2425   & 64.82   & $-$6.64   & $-$18.40(15)      & $-$1.94(14)    & $-$0.17  & $-$13.61   & 9.95   & 8.23    \\
B2020+28     & 68.93   & $-$4.69   & $-$16.8(4)        & $-$9.3(5)      & $-$0.22  & $-$118.62  & 6.46   & 11.91   \\
B2021+51     & 87.93   & 8.36    & 11.1(3)         & 11.3(3)      & 0.29   & 106   & 6.44   & 12.11   \\ \\

B2022+50     & 86.94   & 7.53    & $-$11(3)          & 19(3)        & 0.31   & 209   & 6.37   & 11.99   \\
B2044+15     & 61.18   & $-$16.85  & 0(5)        & 13(6)        & $-$0.70  & 143   & 8.00   & 11.66   \\
B2045$-$16     & 30.58   & $-$33.08  & 42(5)           & $-$107(5)      & $-$0.31  & $-$238  & 6.45   & 12.67   \\
J2051$-$0827   & 39.26   & $-$30.42  & 5(3)            & $-$3.8(17)     & $-$0.53  & $-$16   & 9.75   & 8.38    \\
B2053+21     & 67.90   & $-$14.68  & 5(6)            & $-$3(10)       & $-$0.62  & $-$33   & 6.98   & 12.03   \\ \\

B2053+36     & 79.21   & $-$5.60   & 5(6)            & 5(6)         & $-$0.45  & 109   & 6.98   & 11.46   \\
B2106+44     & 86.98   & $-$2.03   & 8.0(14)         & $-$1.3(14)     & $-$0.18  & $-$31   & 7.88   & 11.28   \\
B2110+27     & 75.06   & $-$14.04  & $-$51(3)          & $-$17(3)       & $-$0.49  & $-$159  & 6.86   & 12.26   \\
B2111+46     & 89.08   & $-$1.28   & $-$12(13)         & $-$7(13)       & $-$0.10  & $-$150  & 7.35   & 11.94   \\
J2124$-$3358   & 10.98   & $-$45.44  & $-$40.8(10)       & 19.4(10)     & $-$0.19  & 17    & 9.58   & 8.51    \\

\hline \end{tabular}\end{table*}\begin{table*}
\begin{tabular}{lrrllrrll}\hline
PSR & $l$ & $b$& $\mu_l$ & $\mu_b$ & Z & $V_Z$ & log[$\tau_c$ (yr)]& log[$B_s$ (G)] \\
 & ($^\circ$) & ($^\circ$) & (mas yr$^{-1}$) & (mas yr$^{-1}$) & (kpc) & (km s$^{-1}$) &  &  \\
\hline
J2129$-$5721   & 338.06  & $-$43.56  & $-$4(3)           & $-$4(3)        & $-$0.94  & $-$19   & 9.45   & 8.45    \\
J2145$-$0750   & 47.84   & $-$42.09  & $-$9.6(7)         & 12.1(4)      & $-$0.34  & 21    & 9.93   & 8.85    \\
B2148+63     & 104.33  & 7.40    & 19(4)           & $-$1(4)     & 0.71   & $-$36   & 7.55   & 11.41   \\
B2148+52     & 97.59   & $-$0.93   & $-$9(3)           & 0(3)      & $-$0.07  & 7     & 5.72   & 12.27   \\
B2154+40     & 90.56   & $-$11.36  & 19.9(9)         & $-$8.4(10)     & $-$0.74  & $-$147  & 6.85   & 12.37   \\ \\

B2217+47     & 98.46   & $-$7.61   & $-$23(8)          & $-$18(7)       & $-$0.29  & $-$188  & 6.49   & 12.09   \\
B2224+65     & 108.71  & 6.83    & 185(3)          & 19(3)        & 0.22   & 166   & 6.05   & 12.41   \\
J2229+2643   & 87.77   & $-$26.30  & $-$5(4)           & $-$13(4)       & $-$0.64  & $-$80   & 10.51  & 7.82    \\
B2227+61     & 107.23  & 3.63    & $-$2(11)          & 10(11)       & 0.26   & 191   & 6.49   & 12.00   \\
J2235+1506   & 80.95   & $-$36.46  & 21(6)           & 1(7)      & $-$0.64  & 2     & 9.78   & 9.49    \\ \\

B2255+58     & 108.90  & $-$0.59   & $-$17(5)          & $-$3(5)        & $-$0.05  & $-$64   & 6.00   & 12.17   \\
B2303+30     & 97.79   & $-$26.67  & $-$4(3)           & $-$17(3)       & $-$1.64  & $-$264  & 6.94   & 12.33   \\
B2310+42     & 104.48  & $-$16.44  & $-$20.7(11)       & 13.6(11)     & $-$0.35  & 77    & 7.69   & 11.30   \\
J2317+1439   & 91.43   & $-$42.38  & 5(2)         & 9(3)         & $-$0.56  & 26    & 10.35  & 7.97    \\
J2322+2057   & 96.59   & $-$37.32  & $-$22(3)          & $-$6(3)        & $-$0.49  & $-$18   & 9.89   & 8.34    \\ \\

B2324+60     & 113.02  & $-$0.01   & 18(5)           & $-$7(5)        & $-$0.00  & $-$161  & 7.02   & 11.46   \\
B2327$-$20     & 49.44   & $-$70.20  & 37(3)           & $-$60(3)       & $-$0.37  & $-$38   & 6.75   & 12.45   \\
B2334+61     & 114.36  & 0.22    & 0(13)  & $-$10(13)      & 0.01   & $-$149  & 4.61   & 12.99   \\
B2351+61     & 116.31  & $-$0.21   & 25(3)           & 1.5(21)      & $-$0.01  & 24    & 5.96   & 12.60   \\
\hline \end{tabular}\end{table*}

  Our sample includes 12 pulsars that are situated within globular
  clusters and for which a 2-D speed can be determined. The
  mean 2-D speed for this subsample is 149(8)\,km\,s$^{-1}$.
  However, this includes the proper motion of the globular clusters as
  well as the pulsars themselves.  For 47~Tucanae, removing the
  \emph{Hipparcos} measurement of the cluster proper motion of
  $\mu_\alpha = 5.3(6)$\,mas\,yr$^{-1}$ and $\mu_\delta =
  -3.3(6)$\,mas\,yr$^{-1}$ (Odenkirchen et al.~1997\nocite{obgt97})
  from the measured proper motions indicates that the intrinsic mean
  2-D speed of the 11 pulsars with measured proper motions in
  the cluster is 25(5)\,km\,s$^{-1}$. That the mean velocity of
  recycled pulsars in globular clusters is significantly lower than
  those moving in the Galactic potential is not surprising given the
  smaller escape velocity of the clusters compared to the Galaxy. For
  47\,Tucanae, for example, the escape velocity is $\sim
  58$\,km\,s$^{-1}$\cite{web85}.  As the measured proper motion of a
  pulsar in a globular cluster is dominated by the cluster motion we
  will leave such pulsars out of the remaining discussion and analysis
  of this paper.

  In Table~\ref{tb:gmotion} we tabulate for each pulsar: Galactic
  position, proper motion in Galactic latitude and longitude, height
  above the Galactic plane, speed perpendicular to the Galactic plane, 
  characteristic age, $\tau_c = P/2\dot{P}$ and surface magnetic
  field strength, $B_s = 3.2\times 10^{19}(P\dot{P})^{1/2}$\,G.  The
  effects of stellar motion and Galactic rotation have been removed
  from the proper motions given in this table.  To do this, we assumed
  a flat Galactic rotation curve with
  galactocentric distance of the Sun, $R_\odot = 8.5$\,kpc and the
  Galactic rotation of the Sun, $V_\odot = 225$\,km\,s$^{-1}$.  These
  are identical to the values used in the analysis by Harrison, Lyne
  \& Anderson (1993) and are similar to the current IAU standard of
  $R_\odot = 8.5$\,kpc and $V_\odot = 220$\,km\,s$^{-1}$ (Kerr \&
  Lynden-Bell 1986).\nocite{klb86}
  Most suggested rotation curves of our Galaxy, e.g.~Olling \&
  Merrifield (1998)\nocite{om98}, only deviate significantly from a
  flat rotation curves for distances from the Galactic centre
  $<3$\,kpc. Since none of the pulsars in our sample are within this region
  of the Galaxy, our assumption 
  of a flat rotation curve does not significantly 
  affect our results.

  Similarly, we choose the same values for the peculiar motion of the
  Sun as Harrison, Lyne \& Anderson (1993) of 15.6\,km\,s$^{-1}$ in
  the direction of $l = 48.8^\circ$ and $b = 26.3^\circ$.  This motion
  is close to the value of 13.4\,km\,s$^{-1}$ obtained by Dehnen \&
  Binney (1997) using \emph{Hipparcos} data.  Software to carry out
  these conversions between equatorial (or ecliptic) and Galactic
  proper motions is available as part of the ATNF pulsar catalogue
  software package (Manchester et al.~2005).  Since the
  uncertainties in pulsar distances discussed in the next section are,
  on average, much larger than the effects of assuming a specific
  rotation curve and solar motion values, the resulting 1-D and 2-D
  speed distributions are unaffected by this correction. In the
  following, we therefore choose not to correct the samples for these
  small effects.

\section{Pulsar distances}\label{sec:distances}

    \begin{table}
    \caption{Pulsars with parallax measurements.  This table contains
    each pulsar's name, parallax, distance as measured from the
    parallax, a reference for the measurement and the pulsar's
    2-D speed assuming the proper motion given in
    Table~\ref{tb:pmsample} and the distance determination.}\label{tb:px}
    \begin{center}\begin{tabular}{lllcl}\hline
   PSR & Parallax & Dist. & Ref. & $V_T$\\ 
       & (mas) & (kpc) &  & (km\,s$^{-1}$) \\ \hline
   B0329$+$54    &   0.94(11) & 1.06(12) &  1 & 98(11)\\
   J0437$-$4715  &   6.98(19) & 0.143(4) &  2 & 94(3)\\
   J0633$+$1746  &   6.4(18)  & 0.16(5) &  3 & 125(40)\\
   B0656$+$14    &   3.47(36) & 0.29(3) &  4 & 59(7)\\
   B0809$+$74    &   2.31(4)  & 0.433(7)&  1 & 103(2)\\ \\
   B0823$+$26    &   2.8(6)   & 0.36(8) &  5 & 186(42)\\
   B0833$-$45    &   3.4(2)   & 0.29(2) &  6 & 80(5)\\
   B0919$+$06    &   0.83(13) & 1.20(19)&  7 & 503(80)\\
   B0950$+$08    &   3.82(7)  & 0.262(5)&  1 & 36(1)\\
   B1133$+$16    &   2.80(16) & 0.36(2) &  1 & 641(36)\\ \\
   B1237$+$25    &   1.16(8)  & 0.86(6) &  1 & 481(33)\\
   B1257$+$12    &   1.3(4)   & 0.8(2)  &  8 & 344(90)\\
   B1451$-$68    &   2.2(3)   & 0.45(6) &  9 & 88(12)\\
   J1713$+$0747  &   0.9(3)   & 1.1(4)  &  10 & 34(13)\\
   J1744$-$1134  &   2.8(3)   & 0.36(4) &  11 & 36(4)\\ \\
   B1855$+$09    &   1.1(3)   & 0.9(2)  &  12 & 27(6)\\
   B1857$-$26    &   $<1.1$   & $>0.9$  &  13 & $>$219\\
   B1929$+$10    &   3.02(9)  & 0.331(10) &  1 & 177(5)\\
   B2016$+$28    &   1.03(10) & 0.97(9) &  1 & 31(3)\\
   B2020$+$28    &   0.37(12) & 2.7(9)  &  1 & 310(100)\\ \\
   B2021$+$51    &   0.50(7)  & 2.0(3)  &  1 & 120(18)\\
   J2145$-$0750  &   2.0(6)   & 0.5(2)  &  14 & 33(9)\\ 
  \hline\end{tabular}\end{center}
  References: 1. Brisken et al.~(2002)\nocite{bbgt02},
  2. van Straten et al.~(2001)\nocite{vbb+01}, 
  3. Caraveo et al.~(1996)\nocite{cbmt96}, 
  4. Brisken et al.~(2003b)\nocite{btgg03b},
  5. Gwinn et al.~(1986)\nocite{gtwr86}
  6. Dodson et al.~(2003)\nocite{dlrm03}
  7. Chatterjee et al.~(2001)\nocite{ccl+01}
  8. Wolszczan et al.~(2000)\nocite{wdk+00}
  9. Bailes et al.~(1990b)\nocite{bmk+90a}
  10. Camilo, Foster \& Wolszczan (1994)\nocite{cfw94}
  11. Toscano et al.~(1999a)\nocite{tbm+99}
  12. Kaspi, Taylor \& Ryba (1994)\nocite{ktr94}
  13. Fomalont et al.~(1999)\nocite{fgbc99}
  14. L{\"o}hmer et al.~(2004)\nocite{lkd+04} 
 \end{table}

  Model-independent distances to 22 pulsars within our sample are
  available from measurements of annual parallax. These distances are
  known to within 10\% accuracy and are listed in
  Table~\ref{tb:px}. Limits on the
  distances for a further 13 pulsars 
  have been determined using HI absorption or by
  associations with globular clusters or supernova remnants (such
  limits are reviewed by Frail \& Weisberg 1990\nocite{fw90}).  The
  distances to the remaining pulsars are estimated from their
  dispersion measures and a Galactic electron density model.  The TC93
  model attempted to predict distances accurate to 25\% or better on
  average. The CL02 model purports to provide distances with a similar
  accuracy (Cordes \& Lazio 2002b\nocite{cl03} report that typical
  distance uncertainties are slightly less than 20\%).  Extensive
  comparisons between the two models have been made (see, for example,
  Cordes \& Lazio 2002b and Kramer et al.~2003\nocite{kbm+03}).  Here,
  we emphasise that the two models are not entirely consistent and
  list major discrepancies for the pulsars considered in this paper in
  Table~\ref{tb:distDisc}. With the CL02 model, only a lower limit of
  50\,kpc is provided for the distance to PSR~B1620$-$09 (the TC93
  model places this pulsar at 3.9\,kpc). The CL02 distance is clearly
  incorrect and hence, we do not include this pulsar in the remaining
  analyses presented in this paper.

  \begin{table}
   \caption{The ten largest distance changes in our sample between the
   TC93 and CL02 models.}\label{tb:distDisc}
   \begin{center}
   \begin{tabular}{lll} \hline
    PSR & D$^{TC}$ & D$^{CL}$  \\
        & (kpc)    & (kpc)     \\ \hline
    B0523$+$11 & $>$7.7  & 3.1 \\ 
    B0559$-$05 & $>$7.5  & 3.9 \\
    B0736$-$40 & $>$11.0 & 2.6 \\
    B0835$-$41 & 4.2     & 1.1 \\ 
    B1552$-$31 & $>$6.1  & 2.3 \\ \\
    B1737$+$13 & $>$4.8  & 1.5 \\
    B1802$-$07 & 11.6    & 5.0 \\
    B1900$-$06 & 8.8     & 5.4 \\
    B2011$+$38 & 13.1    & 8.4 \\
    B2148$+$63 & $>$13.7 & 5.5 \\
   \hline
   \end{tabular}\end{center}
  \end{table}

\begin{figure*}
   \begin{center}
   \includegraphics[width=8cm,angle=-90]{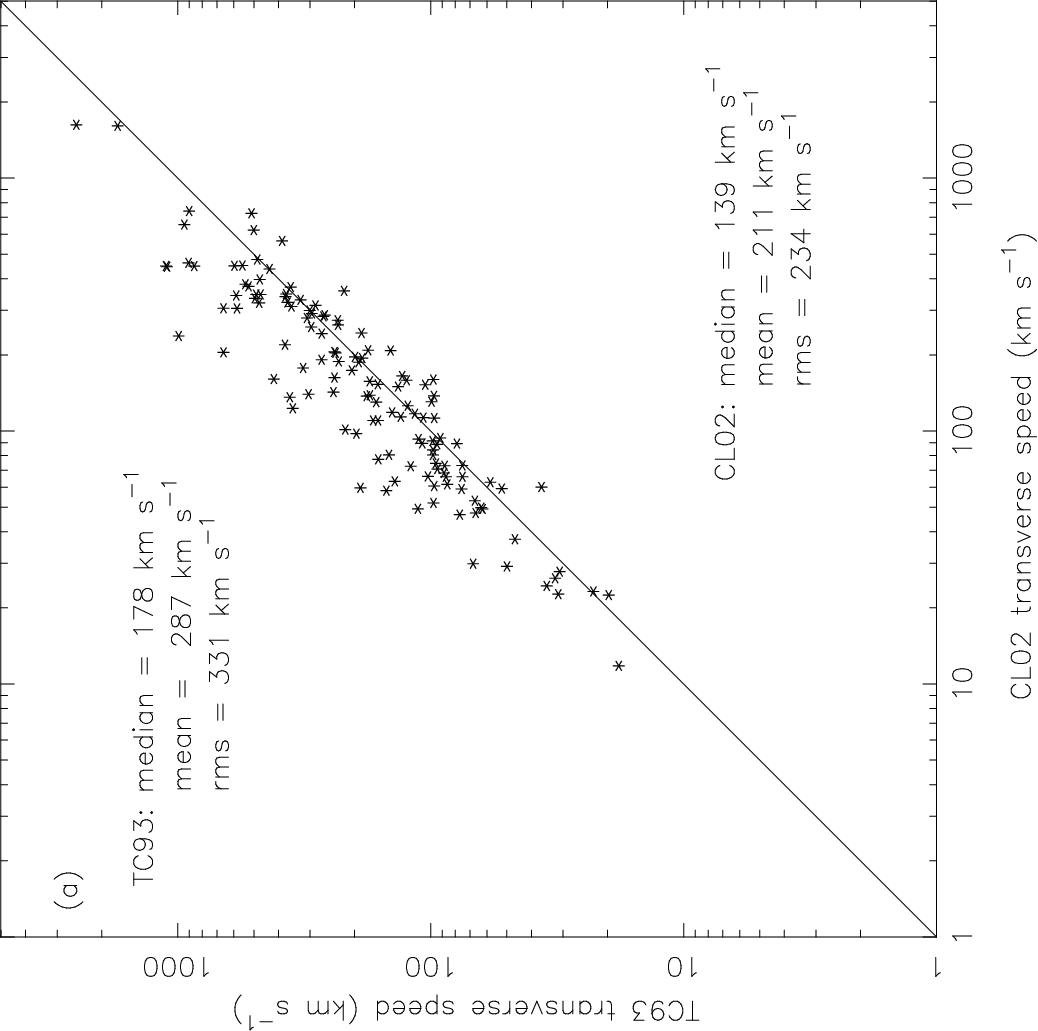}
   \includegraphics[width=8cm,angle=-90]{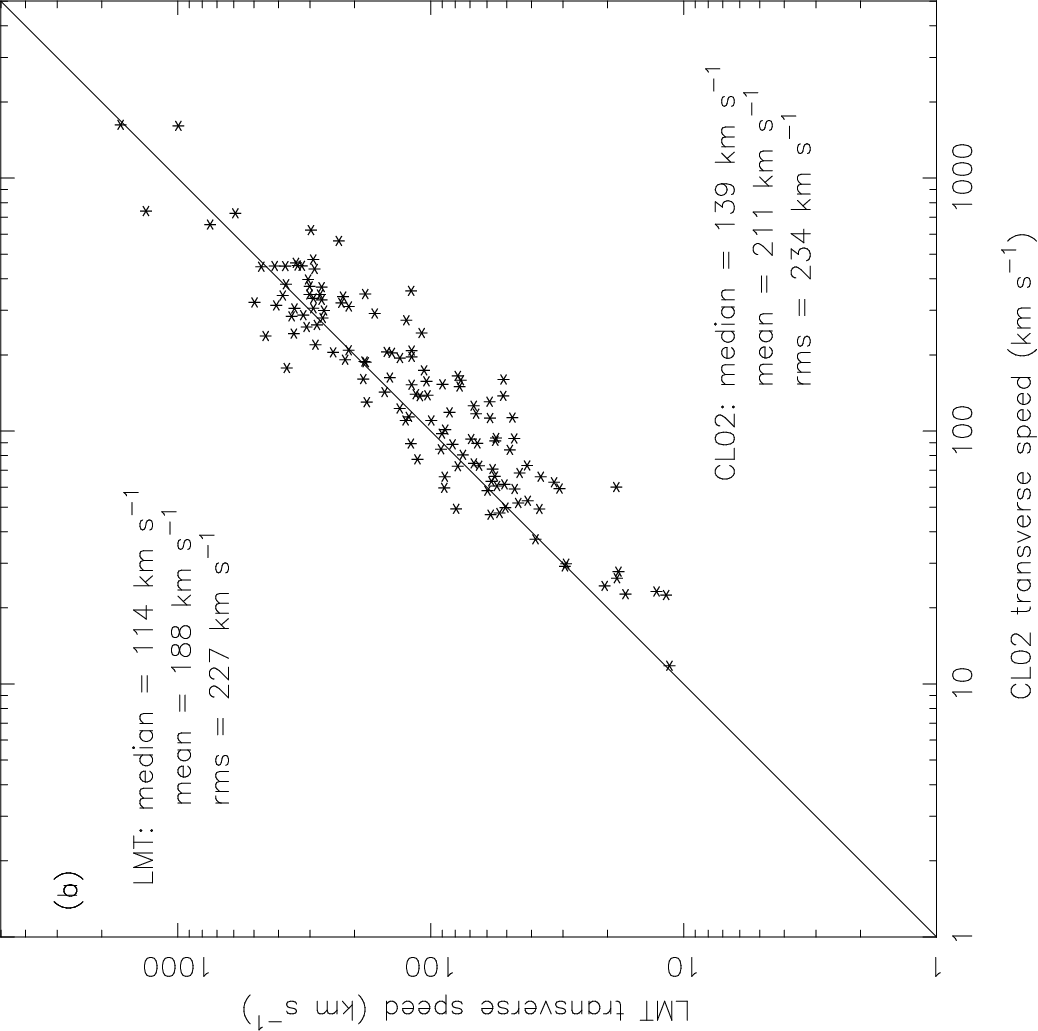}
   \caption{Comparison between pulsar 2-D speeds obtained
   using the CL02 model with: (a) the TC93 model;  (b) the Lyne,
   Manchester \& Taylor (1985) model.}\label{fg:diffDist}
   \end{center}
  \end{figure*}

  As the distances predicted by the CL02 electron density 
  model are, on average,
  smaller than the TC93 model, the derived 2-D speeds are also lower
  (Figure~\ref{fg:diffDist}a).  For our sample, the mean distance has
  decreased by $\sim$2\% (the median decrease is $\sim$10\%). According to
  the TC93 model, four pulsars have 2-D speeds
  $V_T^{TC} > 1000$\,km\,s$^{-1}$, whereas, using the CL02
  model, only two pulsars (PSRs~B2011$+$38 and B2224$+$65) have such
  large 2-D speeds.  The apparent 2-D speeds of the
  remaining two, PSRs~B0523$+$11 and B2148$+$63, have decreased to
  $V_T^{CL} = 446$ and 449\,km\,s$^{-1}$ respectively.

  In passing, we note that the CL02 model places pulsars at distances
  almost halfway between the Lyne, Manchester \& Taylor (1985)
  model\nocite{lmt85} (that was commonly used prior to the TC93 model)
  and the TC93 model (Figure~\ref{fg:diffDist}b).  For the remainder
  of this paper we use the CL02 model (unless independent distance
  estimates exist), but caution that, even with this most recent
  distance model, the distances to some pulsars, and hence
  their velocities, may be significantly under- or over-estimated.  We
  therefore emphasise the importance of continuing to measure pulsar
  distances to refine future electron density models.

\section{One and two-dimensional speeds}

   For all pulsars in our sample at least one component of their
   space velocity is determined. We make use of these by defining a
   pulsar's one-dimensional speed $V_1 = \mu D$,
   where $\mu$ is the proper motion in either the longitudinal or
   latitudinal direction and $D$ is the distance obtained using
   the CL02 model.
   For 156 pulsars we can obtain their 2-D speeds $V_2 = \mu_{tot} D$,
   where $\mu_{\rm tot}$ is the total proper motion.
   In what follows, we investigate aspects of the 
   1-D and 2-D speeds obtained
   for the sample as a whole and in various subsets of the sample. 

  \subsection{Statistical overview}

\begin{figure}
    \includegraphics[angle=-90,width=8cm]{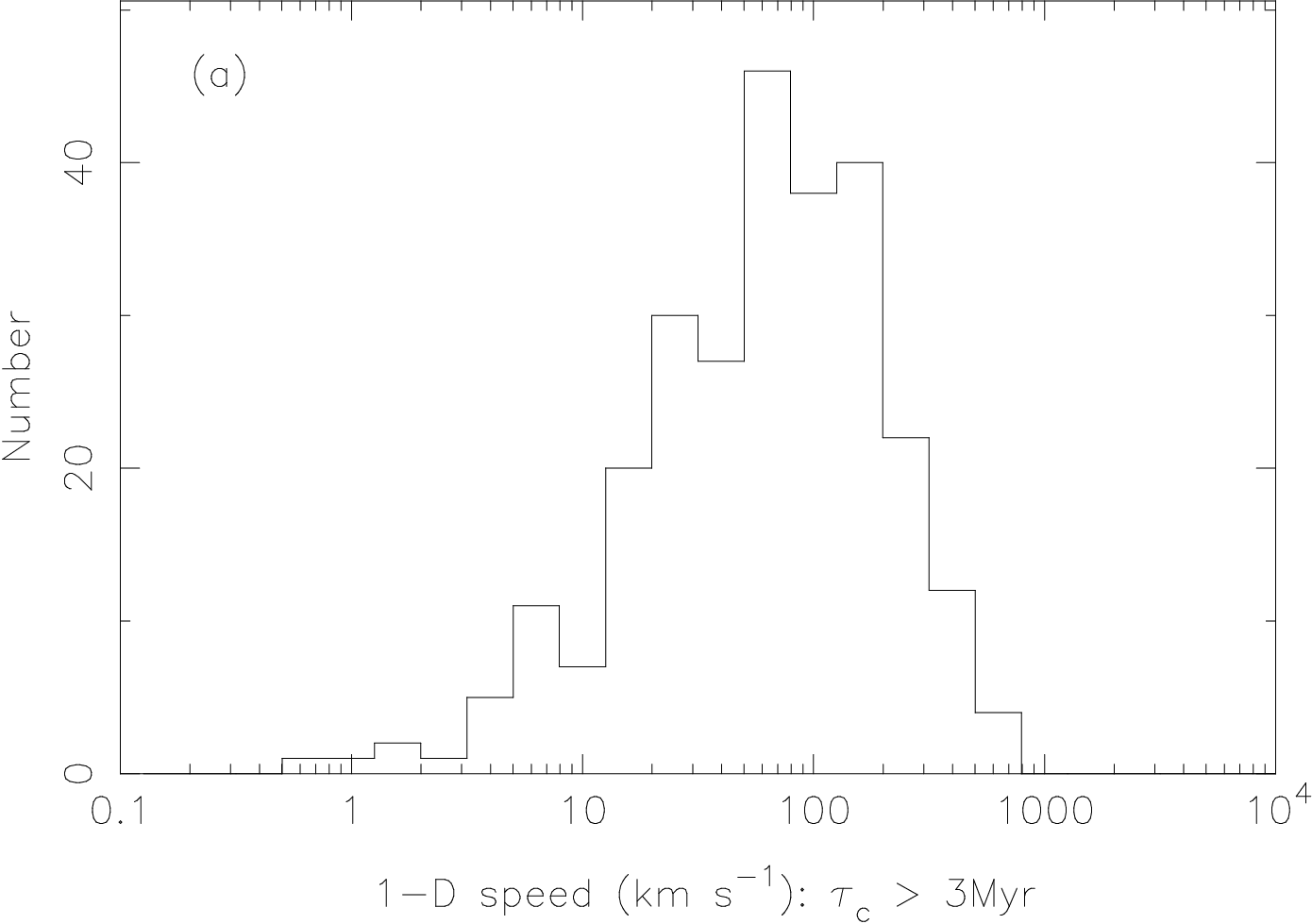} 
    \includegraphics[angle=-90,width=8cm]{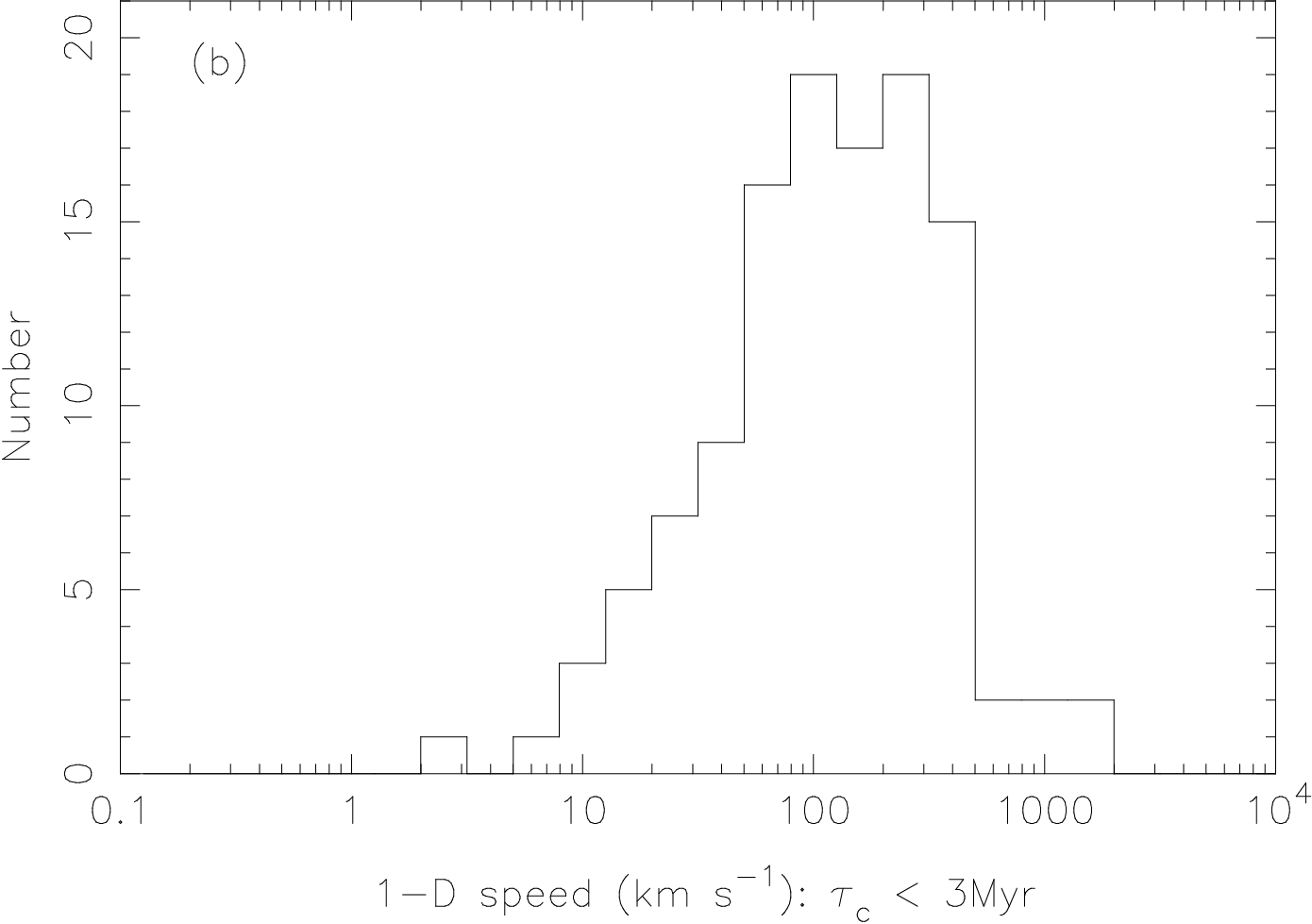} 
    \caption{One dimensional speeds for pulsars in our 
     sample that: a) have characteristic ages greater than 3\,Myr; (b) 
     have characteristic ages less than 3\,Myr.}\label{fg:1dhist}
    \end{figure}

\begin{figure*}
     \includegraphics[angle=-90,width=8cm]{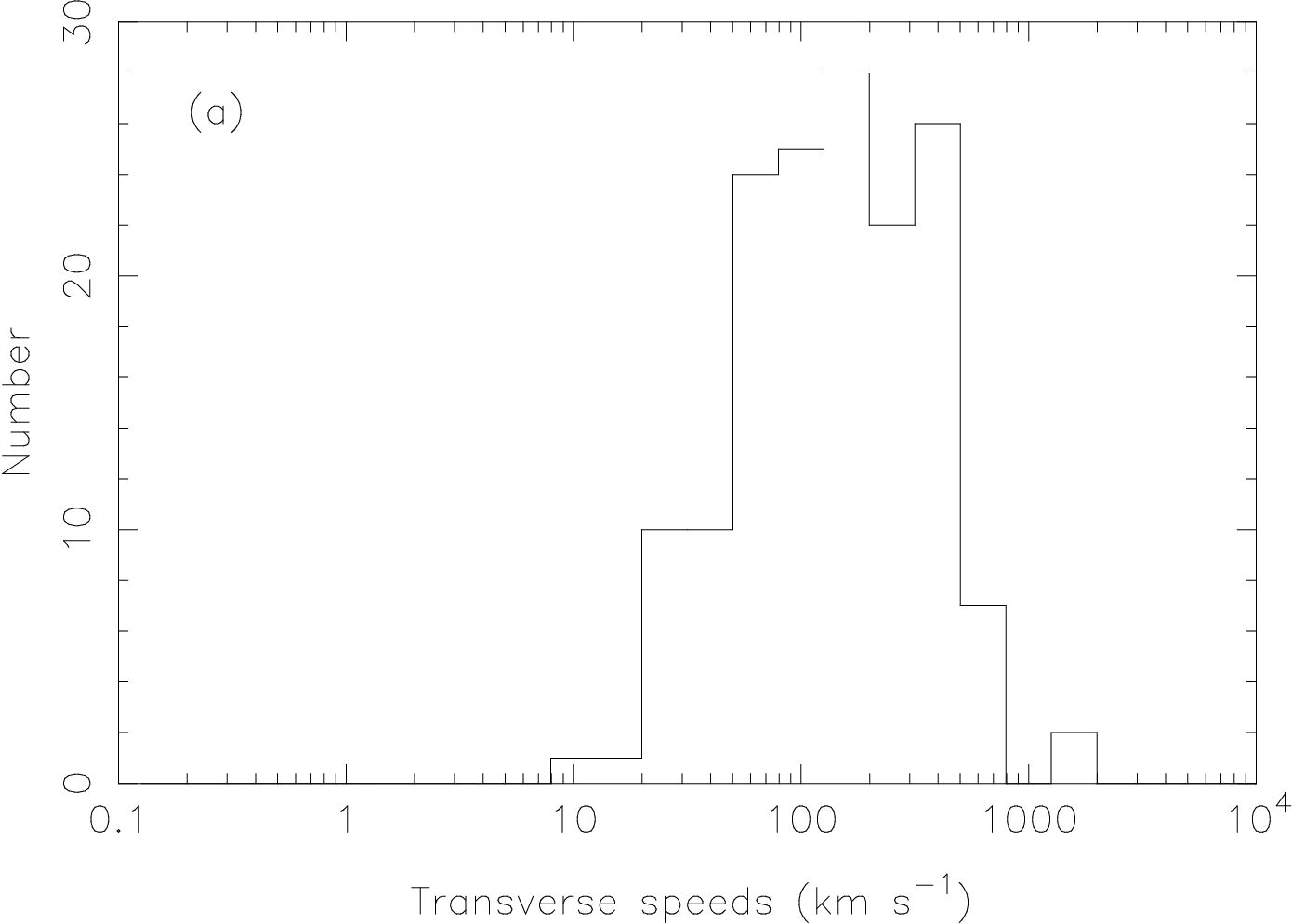} 
     \includegraphics[angle=-90,width=8cm]{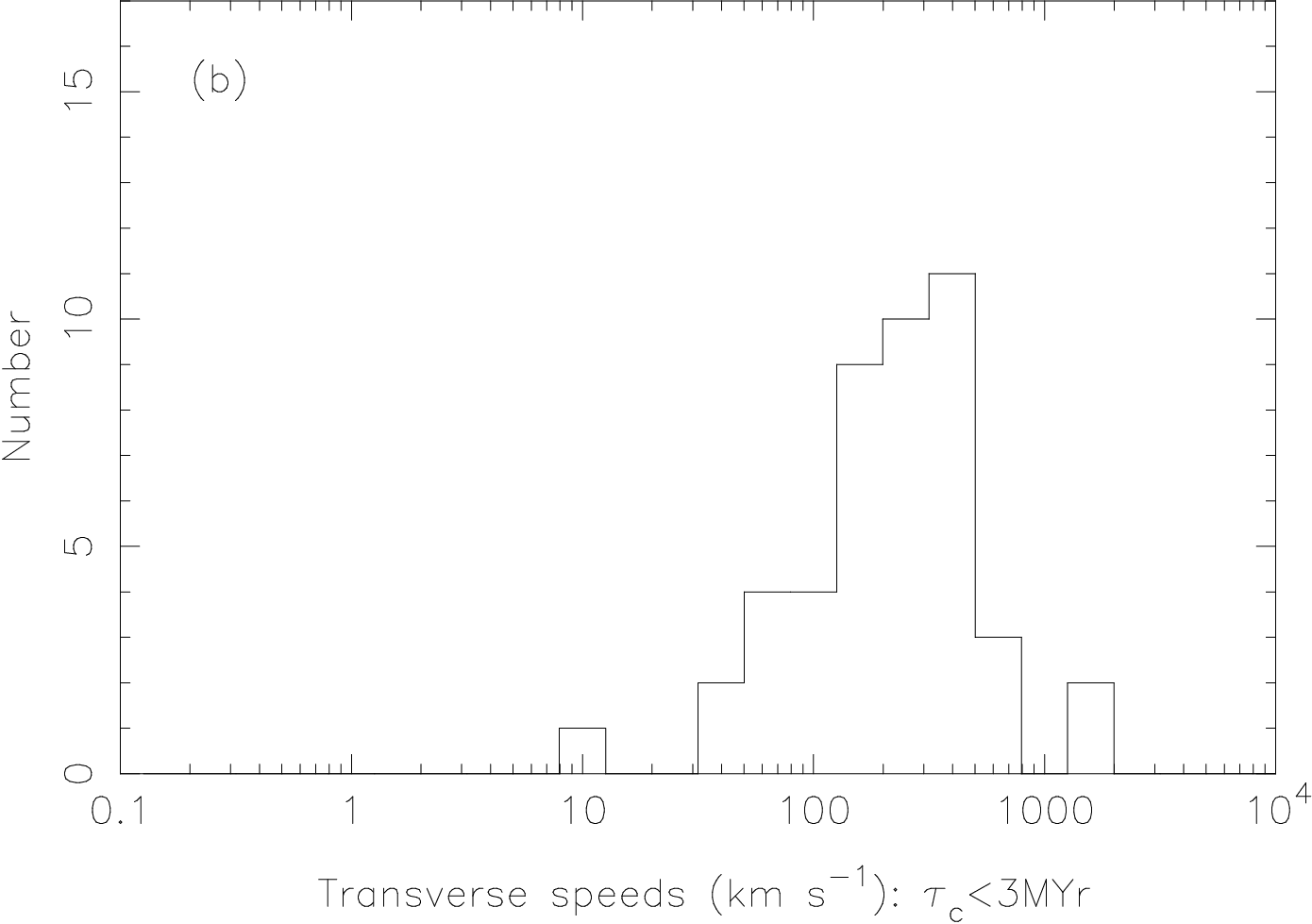}
     \includegraphics[angle=-90,width=8cm]{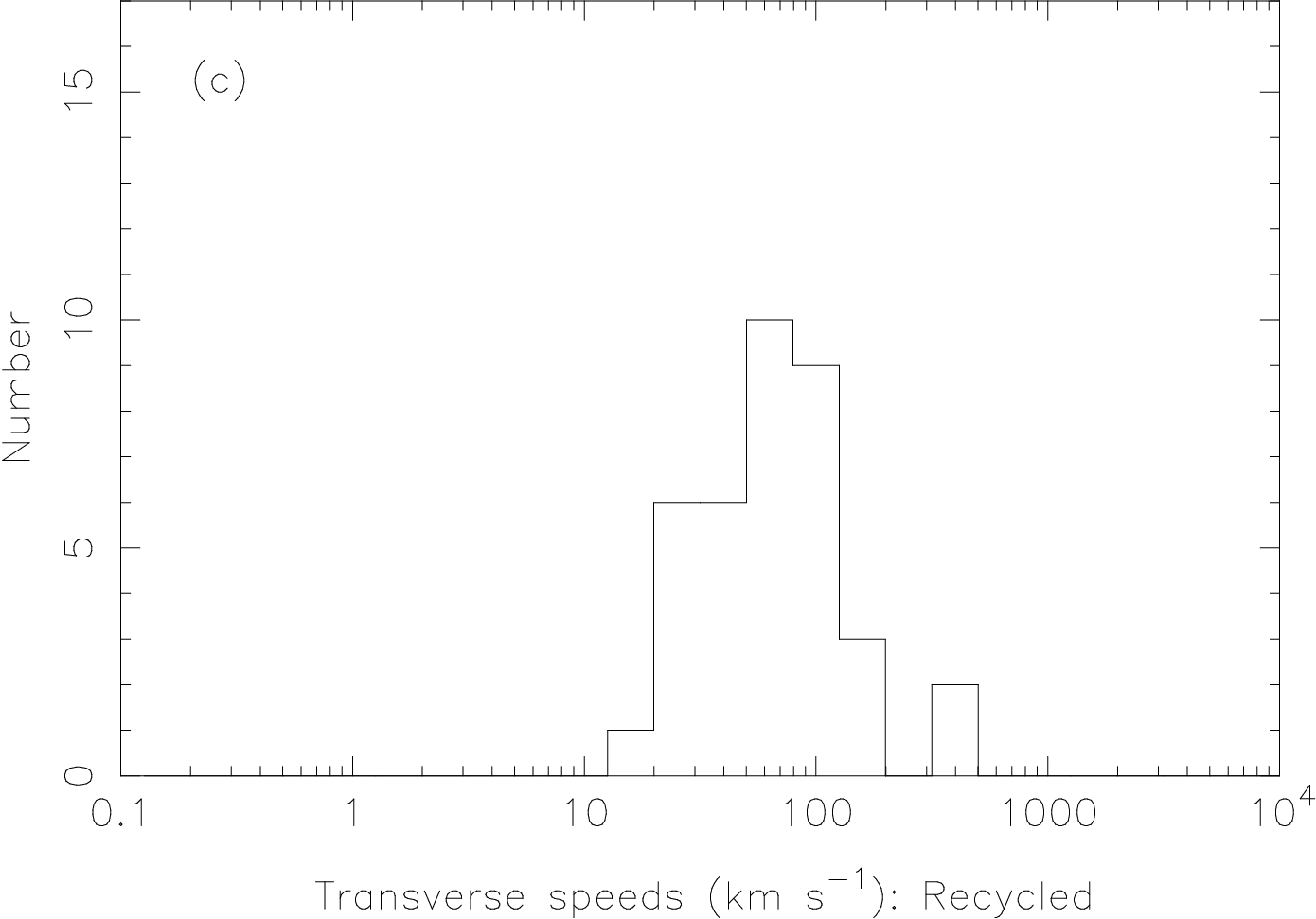}
     \includegraphics[angle=-90,width=8cm]{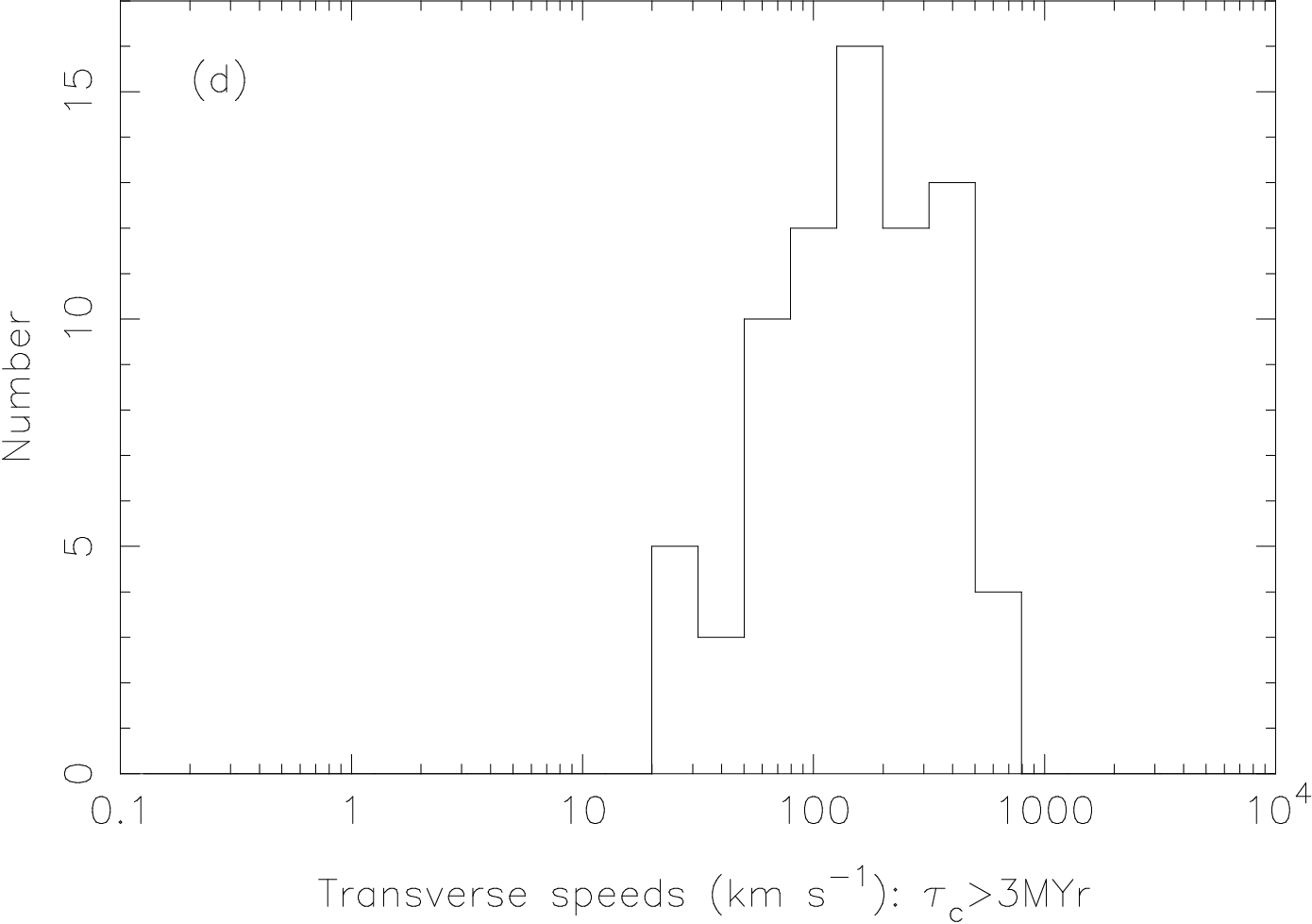}
     \caption{Two-dimensional speeds for: (a) all the pulsars in our
     sample; (b) pulsars with characteristic ages less than 3\,Myr;
     (c) recycled pulsars; (d) normal (not recycled) pulsars
     with characteristic ages greater than 3\,Myr.  The mean 2-D
     speeds for each sample are 211(18), 307(47), 87(13) and
     209(19)\,km\,s$^{-1}$ respectively. Panels b, c and d are drawn
     to the same scale.}\label{fg:2dhist}
    \end{figure*}

   A summary of the 1-D and 2-D speeds for various subsets of the complete
   sample is given in Table~\ref{tb:subset}. Histograms of the data are
   presented in Figures~\ref{fg:1dhist} and \ref{fg:2dhist}. 
   Table~\ref{tb:subset} contains a
   description of each subset, the number of pulsars within this set,
   the number of measured velocity components (for the 1-D speeds, this
   is less than
   twice the number of pulsars as some pulsars have only a proper
   motion measurement in one coordinate)
   and average distances and characteristic ages.

   The mean 1-D speed is 133(8)\,km\,s$^{-1}$ and the fastest moving
   pulsar in our sample, PSR~B2011$+$38, has $V_1=1284$\,km\,s$^{-1}$
   in right ascension and 996\,km\,s$^{-1}$ in declination. The
   smallest upper limit is $V_1<2$\,km\,s$^{-1}$ in right ascension
   for PSR~J1640$+$2224.  The 2-D speeds range from
   12\,km\,s$^{-1}$ for PSR~B1706$-$16 to 1625\,km\,s$^{-1}$ for
   PSR~B2011$+$38.  The mean 2-D speed is
   211(18)\,km\,s$^{-1}$, or 269(25)\,km\,s$^{-1}$ with the TC93
   model.  This latter value is consistent with 300(30)\,km\,s$^{-1}$
   obtained by Lyne \& Lorimer (1994) who also used the same distance model.

   \begin{table*} \caption{Parameters for different samples of
   pulsars. The top half of the table provides values for 1-D speeds,
   the bottom half contains 2-D speeds. N$_{psr}^1$ and N$_{psr}^2$
   indicate the number of pulsars for which 1-D and 2-D speeds have
   been measured. The N$_{psr}^1$ pulsars provide N$_S$ 1-D velocity
   components. The table provides mean and median speeds ($\bar{V}$,
   $V^{med}$), distances ($\bar{D}$, $D^{med}$) and characteristic
   ages ($\bar{\tau}_c$, $\tau_c^{med}$) and an estimate of the mean
   space velocity $V_3$ obtained from the 1-D and 2-D speeds; see text.}\label{tb:subset}
   \begin{center}\begin{tabular}{llllllll} \hline
      & All & Normal & Recycled & SNR & $\tau_c < 3$\,Myr &   $\tau_c > 3$\,Myr & $D < 500$\,pc \\ \hline
   N$_{psr}^1$     & 217 & 178 & 39 & 8  & 73  & 105 & 19 \\
   N$_S$           & 372 & 299 & 73 & 15 & 119 & 180 & 36 \\
   $\bar{V}_1$ (km\,s$^{-1}$) & 133(8) & 152(10) & 54(6) & 150(42) & 192(20) & 126(10) & 77(17) \\
   $V_1^{med}$ (km\,s$^{-1}$) & 79  & 100 & 38 & 95 & 123 & 85  & 54 \\ 
   $\bar{D}$ (kpc) & 2.36(12) & 2.58(14) & 1.32(19) & 2.4(6) & 3.3(3) & 2.06(13) & 0.35(2) \\
   $D^{med}$ (kpc) & 1.91 & 2.14& 0.95 & 2.25 & 2.79 & 1.83 & 0.36 \\
   $\bar{\tau}_c$ (Myr) & 1376 & 70.7& 7333 & 0.1 & 1.2 & 118.9 & 1719 \\
   $\tau_c^{med}$ (Myr) & 7.4 & 4.3& 5610 & 0.07 & 1.1 & 14.1 & 37 \\
   $\bar{V}_3 = 2\bar{V}_1$ (km\,s$^{-1}$) & 266(16) & 304(20) & 108(12) & 300(82) & 384(40) & 252(20) & 154(34) \\ 
   \\		  
   N$_{psr}^2$     & 156 & 121 & 35 & 7 & 46 & 75 & 17 \\
   $\bar{V}_2$ (km\,s$^{-1}$) & 211(18)& 246(22) & 87(13) & 227(85) & 307(47) & 209(19) & 128(33) \\
   $V_2^{med}$ (km\,s$^{-1}$) & 139 & 178 & 73 & 142 & 240 & 152 & 94 \\
   $\bar{D}_2$ (kpc) & 2.25(15) & 2.50(17) & 1.4(2) & 2.7(7) & 3.0(3) & 2.18(17) & 0.35(2) \\
   $D^{med}_2$ (kpc) & 1.70 & 2.07 & 1.04 & 2.5 & 2.10 & 1.99 & 0.36 \\
   $\bar{\tau}_{c,2}$ (Myr)& 1732 & 59.8 & 7511 & 0.06 & 1.1 & 95.7 & 1565 \\
   $\tau_{c,2}^{med}$ (Myr)& 8.7 & 4.82 & 5210 & 0.04 & 0.9 & 10.2 & 18 \\
   $\bar{V}_3 = (4/\pi) \bar{V}_2$ & 269(23) & 313(28) & 111(17) & 289(108) & 391(60) & 266(24) & 163(42)\\
   \hline 
   \end{tabular}
   \end{center}
   \end{table*}

   \subsection{Isotropy of the velocity vector}
   \label{sec:isotropy}

   The observed 1-D and 2-D speeds we measure are, of course,
   the projection of the pulsar's 3-D space velocity. For an
   isotropically distributed pulsar velocity vector of magnitude $V_3$,
   it is straightforward to show that the mean 1-D and 2-D speeds
   are given by $V_3/2$ and $\pi V_3/4$ respectively.
   A simple test of isotropy, then,
   is to compare the ratios of the mean 2-D and 1-D speeds which should
   be $\langle V_2 \rangle/ \langle V_1 \rangle = \pi/2 \simeq 1.57$. 
   For the sample as a whole, we find $\langle V_2 \rangle/ \langle V_1 
   \rangle = 1.6(2)$ in excellent agreement with an isotropic
   velocity vector. Indeed, all the subsamples in Table~\ref{tb:subset}
   give similar results. This simple but important result will be
   used to derive the 3-D velocity distribution in Section 6.

   \subsection{High-velocity pulsars}

  The highest 2-D speeds in our sample are 1624, 1608 and
  740\,km\,s$^{-1}$ for PSRs~B2011$+$38, B2224$+$65 and B1830$-$08
  respectively\footnote{These are the observed 2-D speeds of the
  pulsars.  After removing the effects of Galactic rotation and the
  peculiar motion of the Sun, the 2-D speeds in the pulsar's local
  standard of rest become 1459, 1640 and 730\,km\,s$^{-1}$
  respectively.}.  As there is a large gap between the second and
  third fastest objects, we must consider whether the speeds greater
  than 1000\,km\,s$^{-1}$ are real or due to inaccuracies either in
  the distance model or in the proper motion measurements. If the true
  2-D speed of PSR~B2011$+$38 were 1000\,km\,s$^{-1}$ then the
  observed proper motion would be 25\,mas\,yr$^{-1}$ for its estimated
  distance of 8\,kpc.  This required proper motion is significantly
  lower than recent VLA measurements by Brisken et al.~(2003) who
  obtain $\mu_{tot} = 40(2)$\,mas\,yr$^{-1}$ and from Paper~I where we
  measured $\mu_{tot} = 51(10)$\,mas\,yr$^{-1}$. We can therefore rule
  out the proper motion measurements being in error.  However, as
  there are no independent distance estimates to either PSR~B2011$+$38
  or B2224$+$65 it is possible that the reported distances may be
  overestimated (the reported distance to PSR~B2011$+$38 of 8.4\,kpc
  is the largest in our sample).

  If these 2-D speeds are correct, then it means that the 
  fastest pulsars are still part of a long tail in the velocity
  distribution with the gap between them and the slower-moving pulsars
  being caused by small number statistics
  (i.e.~pulsars at the high end of the distribution are rare and so
  the gap is only an artifact) and do not make up a separate
  population of high-velocity objects.  To investigate whether
  this is possible, we undertook a simple
  simulation where we approximated 
  Figure~\ref{fg:diffDist} by a single 2-D speed distribution.
  In three out of fifteen Monte Carlo realisations of the sample, we
  found a gap above and below 1000\,km\,s$^{-1}$. We therefore conclude that
  the gap observed in Figure~\ref{fg:diffDist} is entirely consistent
  as being the result of small-number statistics of sampling a
  continuous velocity distribution with a large tail.

  To confirm these high 2-D speeds will require independent distance 
  measurements.  However, the observed bow shock for PSR~B2224$+$65 does
  suggest a high velocity for this pulsar which is most likely greater
  than 800\,km\,s$^{-1}$ (see Cordes, Romani \& Lundgren 1993 and
  Chatterjee \& Cordes 2004a\nocite{crl93}\nocite{cc04}).

  \subsection{Young and old pulsars}\label{sec:vel}

  After selecting only those pulsars with characteristic ages less
  than 3\,Myr (Figure~\ref{fg:2dhist}b), we obtain a mean 2-D
  speed of 307(47)\,km\,s$^{-1}$, consistent with
  the value of 345(70)\,km\,s$^{-1}$ obtained by
  Lyne \& Lorimer (1994) who estimated distances using the TC93 model.
  The mean 2-D speed for young pulsars is higher than the
  value of 209(19)\,km\,s$^{-1}$ for older, non-recycled pulsars; see
  Figure~\ref{fg:2dhist}d. However, the velocity distributions for the
  young and older pulsars appear to overlap. We note that the sample
  of 46 young pulsars contains nine objects with 2-D speeds
  less than 100\,km\,s$^{-1}$ and only seven with 2-D speeds
  $>400$\,km\,s$^{-1}$.  The smallest 2-D speed, 
  $12$\,km\,s$^{-1}$, is measured for PSR~B1706$-$16, although whether
  this represents a truly low-velocity population or a projection
  effect is not clear. We investigate this issue further in Section 6.

   \subsection{Pulsar -- supernova remnant associations}

   It is notable that the 1-D and 2-D speeds for the eight pulsars
   that have been associated with supernova remnants (see
   Tables~\ref{tb:subset} and ~\ref{tb:snr}) are significantly lower
   than the mean for other young pulsars and consistent with the mean
   for the entire pulsar sample that includes both young and old
   pulsars.  In contrast, by inferring 2-D speeds based on the offset
   of pulsars from the centre of the supernova remnant, Frail, Goss \&
   Whiteoak (1994) suggested that most pulsars in supernova remnants
   have velocities that are significantly larger than the mean. More
   recent proper motion measurements show that some of the velocities
   claimed by Frail et al.~are too high.  For example, PSR~B1757$-$24
   was thought to be in G5.4$-$1.2, giving rise to a 2-D speed which
   was predicted to be 1800\,km\,s$^{-1}$ but which has subsequently
   been shown to be $\sim$100\,km\,s$^{-1}$, so that it is unlikely to
   be associated with the supernova remnant (Gaensler \& Frail 2000;
   Thorsett, Brisken \& Goss 2002).\nocite{gf00} Other pulsars with
   2-D speeds predicted by Frail et al.~to be greater than
   1000\,km\,s$^{-1}$ are PSRs~B1509$-$58, B1610$-$50, B1800$-$21 and
   B1930$+$22. Unfortunately these pulsars glitch regularly and it is
   difficult to obtain proper motions using timing
   techniques. Interferometric measurements of these pulsars are
   therefore highly desirable.

   One explanation for the observed low velocities of pulsars
   associated with supernova remnants is that any high-velocity
   objects would leave the central regions of the remnant within a
   short timescale and are therefore hard to detect in a targeted
   search of a supernova remnant. The difficulty in identifying a
   genuine association is additionally compounded by the fact that the
   probability of a chance association with a pulsar increases with
   the square of the angular offset from the remnant centre. One
   possible example of a high-velocity pulsar associated with a
   supernova remnant is PSR~B1830$-$08 (nominally the third fastest
   object in our sample with $V_T^{CL} = 740$\,km\,s$^{-1}$) and the
   shell supernova remnant W41 (Clifton \& Lyne 1986; Gaensler \&
   Johnston 1995).\nocite{cl86,gj95a} Regardless of the distance
   uncertainties, the new proper motion measurement for PSR~B1830$-$08
   presented in paper I shows that the pulsar is moving away from W41
   and is consistent with birth within the boundaries of the remnant.
   Further observations to constrain the distance and origin of this
   young pulsar are required.

   \begin{table}
   \begin{center}
   \caption{Pulsars in our sample that have been associated with
   supernova remnants.  Recent work (e.g. Thorsett, Brisken \& Goss 2002)
   indicates that PSR~B1757$-$24 is not associated with G5.4$-$1.2 as
   was previously thought. For PSR~J0538$+$2817 only one component of
   velocity has been determined.}\label{tb:snr}
   \begin{tabular}{llll}\hline
    PSR         & SNR  & $\tau_c$ & $V_T$  \\
                &      & (kyr)  & (km\,s$^{-1}$)  \\ \hline
   B0531$+$21   & Crab          & 1.24 & 140     \\
   J0538$+$2817 & S147          & 618  & $>$140  \\
   B0656$+$14   & Monogem Ring  & 111  & 59      \\
   B0833$-$45   & Vela          & 11.3 & 80      \\
   (B1757$-$24) & (G5.4$-$1.2)  & (15.5) & (89)    \\
   B1830$-$08   & W41           & 41   & 740 \\
   B1951$+$32   & CTB80         & 10.7 & 300     \\
   B2334$+$61   & G114.3$+$0.3  & 40.9 & 159     \\ 
   \hline
   \end{tabular}
   \end{center}
   \end{table}

   In Table~\ref{tb:snr} we include an association between
   PSR~B2334$+$61 and G114.3$+$03 which was first suggested by
   Kulkarni et al.~(1993)\nocite{kpha93} and F{\"u}rst, Reich \&
   Seiradakis (1993).  F{\"u}rst et al.~found that the ages and
   distances of both objects are comparable and that the pulsar is
   located close to the centre of the remnant.  They also noted that,
   assuming the 41 kyr characteristic age to be a good indicator of
   the pulsar's true age, the expected proper motion from birth in the
   centre of the supernova remnant to the pulsar's current position
   should be $\mu_{tot} = 12(2)$\,mas\,yr$^{-1}$.  Our timing
   measurement of $\mu_{tot} =
   11(13)$\,mas\,yr$^{-1}$ is consistent with the predicted
   value for an association. Clearly, given the large uncertainty in
   $\mu_{tot}$, we cannot rule out that the pulsar was not born in the
   centre of the supernova remnant. We can state, however, that
   B2334$+$61 was born somewhere within G114.3$+$0.3 (even with a
   proper motion of $\mu_{tot} = 24$\,mas\,yr$^{-1}$, the pulsar would
   only have moved $\sim$16\,arcmin or only 25\% of the current
   diameter of the supernova remnant within 
   41\,kyr).  The birth position of B2334+61 will be determined
   more accurately with a few more years of timing data or by making
   interferometric measurements.

  \subsection{Recycled pulsars}

    In Table~\ref{tb:msps} we list the 2-D speeds for the
    recycled pulsars in our sample.  The mean value of $V_T =
    87(13)$\,km\,s$^{-1}$ is entirely consistent with the value of
    85(13)\,km\,s$^{-1}$ reported by Toscano et al.~(1999) using the
    TC93 model.  This consistency is due to the known
    millisecond pulsars being relatively close-by and so CL02 and
    TC93 agree much better than they do for more distant pulsars.
    Also, a parallax has been measured for many of these pulsars
    providing independent distance estimates (for the millisecond
    pulsars, only the predicted distances to PSRs~J0218$+$4232,
    J1045$-$4509, J1643$-$1224, J2129$-$5721 and
    J2317$+$1439 have changed by more than 1\,kpc from the TC93 to the
    CL02 model). 

    \begin{table}
    \caption{2-D speeds for the recycled pulsars.  For binary systems,
    the orbital period $P_b$ is also given.  The orbital period listed
    for PSR~B1257$+$12 is that of the most massive planet; see
    text.}\label{tb:msps}
    \begin{center}
    \begin{tabular}{lll} \hline
     PSR & $V_T$ & $P_b$ \\
         & (km\,s$^{-1}$) & (d) \\ \hline
     J0034$-$0534 & 79 & 1.59 \\
     J0218$+$4232 & 63 & 2.03 \\
     J0437$-$4715 & 94 & 5.74\\
     J0613$-$0200 & 59 & 1.19\\
     J0621$+$1002 & 23 & 8.31 \\ \\
     J0711$-$6830 & 89 & --- \\
     J0751$+$1807 & $>$1$^\dagger$ & 0.26 \\
     J1012$+$5307 & 49 & 0.60 \\
     J1022$+$1001 & $>$35$^\dagger$& 7.81 \\
     J1024$-$0719 & 150 & ---   \\ \\
     J1045$-$4509 & 72  & 4.08  \\
     B1257$+$12   & 345 & 66.5 \\
     J1455$-$3330 & 63  & 76.2 \\
     J1518$+$4904 & 28  & 8.63 \\
     B1534$+$12   & 113 & 0.42 \\ \\
     J1603$-$7202 & 47  & 6.31  \\
     J1640$+$2224 & 73  & 176 \\
     J1643$-$1224 & 103 & 147\\
     J1709$+$2313 & 68  & 22.7\\
     J1713$+$0747 & 34  & 67.8\\ \\
     J1730$-$2304 & $>$50$^\dagger$ & --- \\
     J1744$-$1134 & 36  & ---\\
     J1804$-$2717 & $>$7$^\dagger$ & 11.1\\
     B1855$+$09   & 27  & 12.3 \\
     J1909$-$3744 & 80  & 1.53  \\ \\
     J1911$-$1114 & 139 & 2.72  \\
     B1913$+$16   & 73  & 0.32  \\
     B1937$+$21   & 14 & ---  \\
     B1953$+$29   & 84 & 117 \\
     B1957$+$20   & 359& 0.38\\ \\
     J2019$+$2425 & 160& 76.5\\
     J2051$-$0827 & 26 & 0.10 \\
     J2124$-$3358 & 63 & ---  \\
     J2129$-$5721 & 52 & 6.63 \\
     J2145$-$0750 & 33 & 6.84 \\ \\
     J2229$+$2642 & 117& 93.0 \\
     J2235$+$1506 & 91 & ---  \\
     J2317$+$1439 & 31 & 2.46 \\
     J2322$+$2057 & 95 & ---  \\   
    \hline \end{tabular}
    \end{center}
    $\dagger$ For these pulsars, only the proper motion coordinate in
    ecliptic longitude has been measured.  The given value is
    therefore a lower limit to its 2-D speed.
    \end{table}

    Toscano et al.~(1999) also noted that isolated millisecond pulsars
    have lower velocities than binary millisecond pulsars (see also
    Johnston et al.~1998\nocite{jnk98}).  Our sample contains seven
    solitary and 28 binary millisecond pulsars with measured
    2-D speeds. The size of this sample more than doubles that
    of Toscano et al.~(1999). The mean 2-D speeds we obtain for
    the solitary and binary millisecond pulsars are 77(16) and
    89(15)\,km\,s$^{-1}$ respectively which are not significantly
    different.

\begin{figure}
    \includegraphics[angle=-90,width=8.5cm]{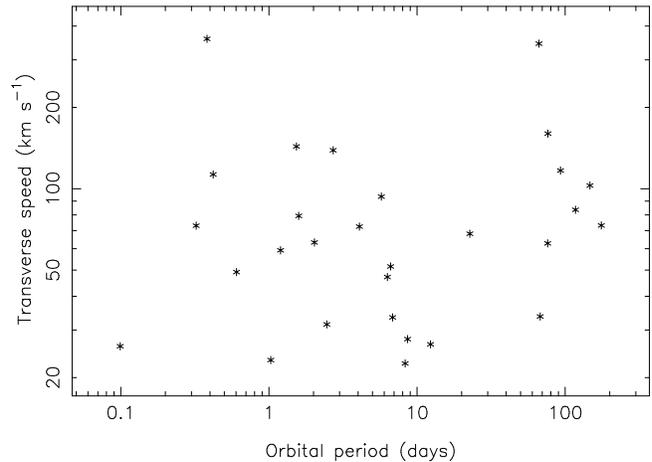} 
    \caption{2-D velocity versus orbital period for
    binary millisecond pulsars.}\label{fg:vtPb}
    \end{figure}

    Tauris \& Bailes (1996)\nocite{tb96} predicted that the velocities
    of binary millisecond pulsars should increase with decreasing
    orbital period. No such correlation was found in their data which
    they attributed to selection effects. Our data, shown in
    Figure~\ref{fg:vtPb}, also show no correlation.  Tauris \&
    Bailes (1996) predict a lower limit for the velocity of such
    pulsars of $\approx 25$\,km\,s$^{-1}$ and an upper limit of
    $\approx$\,270\,km\,s$^{-1}$.  Two pulsars in our sample,
    PSRs~B1257$+$12 and B1957$+$20, with 2-D speeds of
    344 and 359\,km\,s$^{-1}$ respectively, are clearly well above
    this upper limit. The first, PSR~B1257$+$12 is known to be part of
    a planetary system and may have undergone a different evolution
    than other binary millisecond pulsars.  The second, the
    `black-widow' pulsar PSR~B1957$+$20, is thought to be slowly
    ablating its companion \cite{fst88} and may, therefore, provide a
    link between binary and isolated millisecond pulsars.  This seems
    rather unlikely since we do not see
    solitary millisecond pulsars with such high velocities.

    We may expect double neutron star systems to have even lower
    velocities than millisecond pulsars as a consequence of having
    survived two supernova explosions.  Unfortunately, we currently
    only have three double neutron star systems with proper motion
    determinations.  Our proper motion measurement for PSR~J1518+4904
    from Paper~I of 25(7)\,km\,s$^{-1}$ agrees well with
    28(2)\,km\,s$^{-1}$ found by by Nice, Taylor \& Sayer
    (1999)\nocite{nts99} (only the uncertainties in the total proper
    motion were considered when determining the errors on these
    values). Earlier measurements for PSR~B1534+12 and B1913+16 imply
    respective 2-D speeds of 113.0(1) and 73(2)\,km\,s$^{-1}$
    respectively. More recently, from scintillation measurements of
    the double pulsar J0737$-$3039, Coles et al.~(2004)\nocite{cmr+04}
    derive a 2-D speed of $\sim$66\,km\,s$^{-1}$. A proper motion
    measurement of this system should be available soon through both
    interferometric and timing measurements. Given the small-number
    statistics, we simply note that these 2-D speeds are
    entirely consistent with the distribution for the whole sample of
    recycled pulsars.

\section{Motion in the Galactic plane}\label{sec:gplane}

   Standard models of pulsar birth suggest that pulsars are born in
   Type~II supernovae (see, for example, Gunn \& Ostriker 1970).
   The progenitor massive stars lie along the
   Galactic plane with a scale height of $\sim$63\,pc (Sun \& Han
   2004).  Any velocity kick imparted to the pulsar at its
   birth will make it move away from the Galactic plane.  Hence, the
   motion of young pulsars that appear to be moving toward the plane
   can only be explained by projection effects or by the pulsar being
   born well above the plane and out of the expected progenitor
   region.  For instance, the unknown radial motion of a pulsar may
   make it appear to be moving towards the plane when it is, in fact,
   moving away (see, for example, Helfand \& Tademaru 1977).
   Restricting our study to pulsars with characteristic ages less than
   3\,Myr guarantees that all bound objects are within the first
   quarter-cycle of their oscillation perpendicular to the Galactic
   plane (see, for example, Cordes \& Chernoff 1998).

   Harrison, Lyne \& Anderson (1993)\nocite{hla93} found that the
   positions and proper motions of PSRs~B0450$-$18, B0523$+$11,
   B0540$+$23 and B1718$-$02 could only be explained by the pulsars
   being born above the Galactic plane and currently moving moving
   toward the plane. However, Cordes \& Chernoff (1998) analysed the
   proper motions of 49 pulsars with characteristic ages less than
   10\,Myr and found that all were consistent with the earlier
   conclusion of Helfand \& Tademaru (1977) that all pulsars are born
   in the plane and subsequently move away.  The main difference
   between the analyses was that Harrison, Lyne \& Anderson (1993)
   defined a much smaller progenitor region than Cordes \& Chernoff
   (1998). For instance, even though PSR~B0540$+$23 is likely to be
   moving towards the Galactic plane, due to its low height above the
   Galactic plane of only 0.2\,kpc, it was deemed, by Cordes \&
   Chernoff (1998), to have been born within the expected pulsar
   progenitor region of the Galaxy.

   In our sample, only PSR~B0114$+$58 (and B0540$+$23; see above
   discussion) has a characteristic age less than 3\,Myr, a
   current $z$-height greater than 100\,pc and is seemingly moving
   toward the plane. However, this pulsar lies below the plane at a
   distance of $130$\,pc which is within current estimates of the
   pulsar progenitor scale height. We therefore agree with Cordes \&
   Chernoff (1998) that all the pulsars that seem to be moving toward
   the plane are consistent with having been born within the plane.

   \begin{figure}
    \includegraphics[angle=-90,width=8cm]{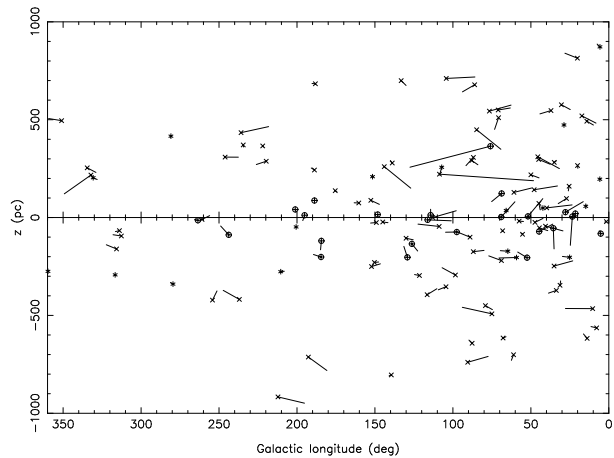}
    \caption{The motion of pulsars with respect to the Galactic
    plane.  Pulsars indicated by a circle have $\tau_c < 1$\,Myr,
    those with $\tau_c > 1$\,Gyr by a `*' symbol.  All other pulsars
    are indicated using an `x'.}
    \label{fg:z_gl}
   \end{figure}

   In Figure~\ref{fg:z_gl} we indicate the angle of velocity vectors
   with respect to the Galactic plane for pulsars with Galactic
   latitudes less than 30$^\circ$. The lengths of the tracks are
   indicative of the magnitude of the pulsar velocities and were
   calculated using the pulsar proper motions in Galactic longitude
   and latitude only; as before, the radial velocity component has
   been ignored.  We find no predominant angle for the motion of the
   pulsars with respect to the Galactic plane.

   Earlier we analysed pulsar velocities based on proper motion
   measurements in celestial or ecliptic coordinates.  The difference
   between these velocities and those obtained after the effects of
   Galactic rotation and the peculiar motion of the Sun have been
   removed is minimal and is smaller than the likely uncertainties due
   to the pulsar distances.  The shape of the 2-D speeds after
   taking into account the effects of Galactic rotation is little
   changed from the earlier version (the mean velocities are
   identical).  However, the velocities for a few pulsars change
   significantly.  For instance, in the most extreme case the velocity
   of PSR~B1933+16 reduces from 346 to 208\,km\,s$^{-1}$.

\section{Three-dimensional velocities of young pulsars}

    With the exception of two binary pulsars
    (Bell et al.~1995; van Kerkwijk et al.~1996)
    \nocite{bbs+95,kbk96} it is generally
    not possible to make a direct measurement of the
    true 3-D space velocity of a pulsar.  However, as the 
    observed data are consistent with an isotropic velocity
    vector, it is possible to determine the 3-D speed distribution
    from the observed 1-D and 2-D speeds.  
    In this section we describe and discuss the consequences of
    a new technique that deconvolves the expected form of a
    single-magnitude isotropic space velocity from the observed 
    1-D and 2-D distributions. 

    \begin{figure*}
     \includegraphics[angle=-90,width=16cm]{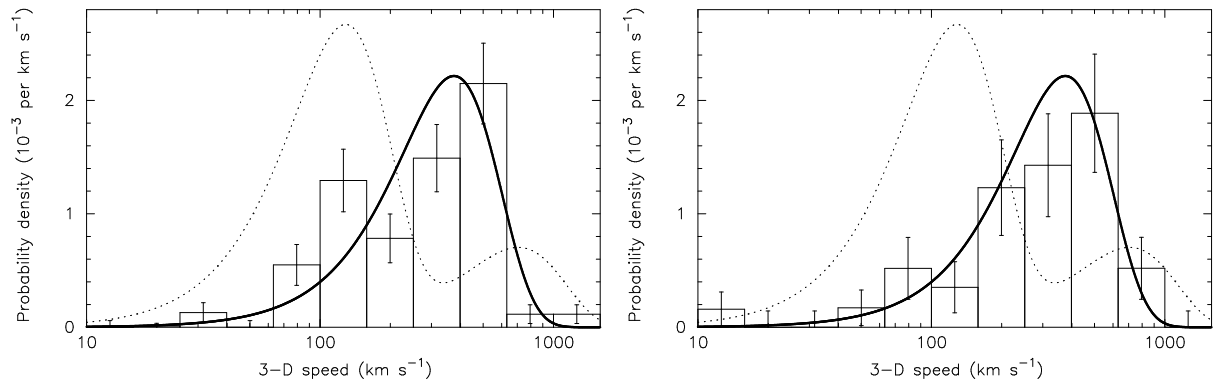}
     \caption{The 3-D velocity distributions obtained from the
     observed 1-D (left) 2-D (right) distributions using the
     deconvolution technique described in the text.  The uncertainties
     on each histogram bin are calculated as $\sqrt{N}$ where $N$ is
     the number of pulsars in each bin.  The histograms are normalised
     to unit area. The dotted curve shows the 3-D
     distribution favoured by Arzoumanian et al.~(2002). The solid
     curve is the best-fitting Maxwellian distribution
     to the histogram from the 2-D distribution with
     $\sigma=265$ km~s$^{-1}$.}
     \label{fg:3D}
    \end{figure*}

 \subsection{The deconvolution technique}

    Following Lyne \& Lorimer (1994), we derive the birth velocity
    distribution using the sample of pulsars with $\tau_c<3$ Myr to
    minimise any selection biases against detecting older
    high-velocity pulsars.  The resulting samples of 119
    1-D and 46 2-D
    speeds therefore represent two projected views of the
    underlying 3-D distribution which we seek to recover.
    For a single-magnitude isotropic
    space velocity distribution of $V_3$, the probability density
    function corresponding to a logarithmic bin $V_1$ for 1-D speeds
    \begin{eqnarray}
     P(V_1)d \log(V_1) = \frac{V_1}{V_3}d\log(V_1)\\ \nonumber (0 < V_1 < V_3).
    \end{eqnarray}
    Similarly, for the 2-D case we find
    \begin{eqnarray}
     P(V_2)d \log(V_2) = \frac{V_2^2}{V_3}\frac{d\log(V_2)}
     {(V_3^2-V_2^2)^{1/2}} \\ \nonumber (0 < V_2 < V_3).
    \end{eqnarray}
    One way to think about these 1-D and 2-D distributions is that
    they represent the convolution of the underlying 3-D distribution
    with the 1-D and 2-D probability density functions.  We therefore
    can use a CLEAN-type algorithm (Hogb\"om 1974) \nocite{hog74} to
    deconvolve these functions from the observed speed distributions.
    In detail, we (i) cross-correlate the analytic expression with the
    observed distribution to determine the velocity, $V_3$,
    corresponding to the maximum correlation; (ii) subtract a scaled
    version of the analytic expression defined by $V_3$ from the
    observed distribution with a small gain factor; (iii) record a
    ``CLEAN''-component corresponding to $V_3$ and repeat for a large
    number of iterations.  The ``CLEAN''-components are subsequently
    binned to give a histogram representing the true 3-D space
    velocites.

 \subsection{The underlying space velocity distribution}

  The resulting 3-D distributions obtained from the 1-D and the 2-D
  observed distributions are shown in Figure~\ref{fg:3D}. Both forms
  of the 3-D distribution are in excellent agreement with each other
  and have mean 3-D speeds of 400(40) and 431(60)\,km\,s$^{-1}$
  respectively. The smaller error on the mean derived from the 1-D
  speeds reflects the larger size of this sample.  It is notable that
  the long ``tail'' towards low speeds in the observed 1-D histogram
  (Figure~\ref{fg:1dhist}b) is not present in the 3-D histogram. This
  reflects the fact that the observed ``tail'' is simply due to
  projection effects.  Our 3-D distributions show that very few young
  pulsars have 3-D speeds less than $\sim$60\,km\,s$^{-1}$.  The right
  panel of Figure~\ref{fg:3D} shows that an excellent fit (reduced
  $\chi^2=0.6$) can be obtained using a simple Maxwellian distribution
  with a 1-D rms $\sigma=265$~km~s$^{-1}$.

  We have tested our deconvolution algorithm by simulating 3-D
  velocity distributions of different forms, randomly selecting as
  many 1-D speeds as in our real sample from these 3-D distributions,
  binning the resulting 1-D speeds as in Figure~\ref{fg:1dhist} and
  using our deconvolution method to derive the underlying 3-D
  velocities. In all cases, we have successfully reproduced the
  original form of the simulated distribution. It is of particular
  interest to determine whether a single component velocity
  distribution could be obtained using our algorithm from a simulated
  bimodal 3-D distribution.  We have therefore formed a bimodal 3-D
  distribution from two Gaussians of differing amplitudes centred on
  100 and 630\,km\,s$^{-1}$ similar to that proposed by Arzoumanian et
  al. (2002).  Using the technique described above, we attempted to
  derive a 3-D distribution using our deconvolution method.  In all
  cases we reproduced a bimodal distribution.  The zero-lag cross
  correlation between the predicted and the simulated distributions
  provided a measure of the goodness of the reproduced distribution.
  All the correlation coefficients were found to be close to 1
  implying that the deconvolution techqniue accurately reproduces
  the bimodal distribution, the positions of the components and their
  relative amplitudes.  Similar results are obtained when simulating a
  single component distribution.

  \subsection{Comparison with earlier work}

  Although most authors agree that a broad spectrum of velocities is
  required to explain the observations, considerable disagreement
  exists on whether the distribution has more than one component.
  Based on a number of available constraints, Fryer, Burrows \& Benz
  (1998)\nocite{fbb98} favour a bimodal velocity distribution with one
  peak near 0\,km\,s$^{-1}$ and another above
  600\,km\,s$^{-1}$. Similarly, from an extensive study of a sample of
  pulsars detected in low-frequency ($\sim 0.4$~GHz) radio surveys,
  Arzoumanian, Chernoff \& Cordes (2002)\nocite{acc02} find a two
  component distribution with characteristic velocities of 90 and
  500\,km\,s$^{-1}$.  Whether the origin of these two components is
  due to a physical effect of the kick mechanism is still a matter for
  debate.

  The results of our novel deconvolution analysis of a large
  sample of pulsars do not support the idea of a bimodal distribution
  of pulsar velocities. This can be seen 
  in the left panel of Figure~\ref{fg:3D} where we compare our
  results with an appropriately normalised version of
  the 3-D distribution preferred by Arzoumanian et al.~(2002).
  This figure clearly shows that the new distribution is incompatible
  with the Arzoumanian et al.~model which predicts an excess of low-velocity
  pulsars and dearth of high-velocity ones.
  Fryer et al.~(1998) argued that the pulsar proper motion data
  available to them at that time were not sufficient to constrain 
  distribution of neutron star kick velocities. It would be instructive
  (but beyond the scope of this paper)
  to repeat their calculations using the larger sample of proper
  motions now available. For
  now, we re-iterate that a Maxwellian distribution provides 
  an excellent fit to the 3-D pulsar velocity distribution.
  
  An earlier analysis by Hansen \& Phinney (1997)\nocite{hp97}
  on a smaller sample of pulsars
  also preferred a Maxwellian distribution for $V_3$. However, 
  their value of $\sigma=190$~km~s$^{-1}$ does not provide a 
  very good fit (reduced $\chi^2=5.7$) to the new larger sample
  presented here. Our results, however, do agree well with the mean
  velocity of 450(90) km~s$^{-1}$ found by Lyne \& Lorimer (1994).
  The larger  sample of pulsars in this analysis has resulted in a reduction of
  the uncertainties by more than a factor of two.
  The slight reduction in the mean value of 400(40) km~s$^{-1}$ compared
  with Lyne \& Lorimer (1994) is largely a result of the smaller
  average distances inferred from the CL02 model compared with TC93.

  The different pulsar velocity distributions proposed 
  predict different fractions of pulsars that will escape the
  Galactic gravitational potential. Assuming the escape velocity of
  our Galaxy to be $\sim 430$\,km\,s$^{-1}$ \nocite{lt90}(Leonard \&
  Tremaine 1990), we find that 35\% of all pulsars will escape the
  Galactic potential in the new distribution compared to 50\% for the
  bimodal distribution of Arzoumanian et al.~(2002) and only 16\% for
  the Hansen \& Phinney (1997) distribution.

  In summary, our new results on the 3-D pulsar velocity distribution
  more closely match the earlier conclusions of
  Lyne \& Lorimer (1994) than the results of Hansen \& Phinney (1997),
  Fryer, Benz \& Burrows (1998) or Arzoumanian et al.~(2002).  
  While we have tried to account for
  selection effects by placing a 3~Myr cut-off in our sample of young
  pulsars, a more detailed analysis that fully takes into account the
  various selection effects present in this new sample (and ideally
  including the other constraints discussed by Fryer et al.~1988) should now be
  carried out to further investigate the results presented here.

\section{Conclusions}

   We have presented an updated catalogue of the kinematics of 233
   pulsars, effectively doubling the size of the previous sample.
   Our results may be summarised as follows:

   \begin{enumerate}

     \item{In contrast to the results of Frail, Goss \& Whiteoak
    (1994), most pulsars within supernova remnants are found to have
    lower velocities than other pulsars.  This is most likely due to
    selection effects where fast moving pulsars leave the supernova
    shell within a relatively short period of time.}

    \item{The proper motions for PSRs~B1830$-$08 and B2334$+$61 are
    consistent with their proposed associations with the supernova
    remnants W41 and G114.3+0.3.}

    \item{The fastest moving pulsar with a well-defined distance is
    PSR~B1133$+$16 which has a 2-D speed of
    640\,km\,s$^{-1}$.  However, according to the CL02 (and TC03)
    distance model PSRs~B2011$+$38 and B2224$+$65 both have 2-D
    speeds greater than 1500\,km\,s$^{-1}$.}
 
    \item{The CL02 distance model generally predicts smaller
    distances, and hence 2-D speeds, than the TC03 model.  The mean
    1-D and 2-D speeds for pulsars with characteristic ages less than
    3\,Myr are 192(20) and 307(47) km\,s$^{-1}$. The observed
    1-D and 2-D speeds clearly demonstrate that the 3-D velocity vector
    is isotropic.}
      
    \item{Based on a deconvolution analysis of the new samples of 1-D
    and 2-D speeds of young pulsars, we find the mean 3-D birth speed
    to be 400(40)\,km\,s$^{-1}$. The 3-D speeds are well fit by a
    Maxwellian distribution with 1-D rms $\sigma=265$~km~s$^{-1}$. We
    find no evidence for a bimodal velocity distribution.}
   
   \end{enumerate}

    The implications of these results for `kick' mechanisms may be
    summarised by stating that the true space velocities of
    young pulsars range from a few 10s to well over 1000\,km\,s$^{-1}$
    with a mean velocity of 400(40)\,km\,s$^{-1}$.  According to
    Lai, Chernoff \& Cordes (2001): (1) local convective
    instabilities in the collapsed stellar core can account for
    velocities up to $\sim 100$\,km\,s$^{-1}$; (2) global asymmetric
    perturbations can create velocities over 1000\,km\,s$^{-1}$; 
    (3) asymmetric neutrino emission can provide kick velocities up
    to $\sim 1000$\,km\,s$^{-1}$; (4) the electromagnetic rocket effect
    can accelerate pulsars up to similarly high velocities.  Our
    results suggest that (1) is unlikely.  (4) is testable by
    studying the alignment between the direction of motion of a pulsar
    and its spin-axis (see Deshpande, Ramachandran \& Radhakrishnan
    1999)\nocite{drr99} although the duration of the kick will also
    affect the observed alignment \cite{sp98}. 

   More proper motions will become available within the next few years
   both from interferometry and from timing.  The Jodrell Bank data
   archive will be able to provide values or limits on the proper
   motions of many hundreds of pulsars in the near future. These
   will further improve the constraints on the distribution and origin
   of pulsar velocities.

\section*{Acknowledgements}

  During the course of this work we made
  extensive use of NASA's Astrophysics Data System bibliographic database
  and the astro-ph preprint service.
  DRL is a University Research Fellow funded by the Royal Society.

\bibliography{modrefs,psrrefs,crossrefs}
\bibliographystyle{mn}

\end{document}